\documentclass[11pt]{article}

\usepackage[margin=1.1in]{geometry}
\usepackage[T1]{fontenc}
\usepackage{lmodern} 
\usepackage{amsmath,amssymb,amsthm,mathtools}
\usepackage{physics}
\usepackage{graphicx}
\usepackage{enumitem}
\usepackage{microtype}
\usepackage{authblk}
\usepackage[colorlinks=true,allcolors=magenta]{hyperref}
\usepackage{cleveref}

\theoremstyle{plain}
\newtheorem{theorem}{Theorem}[section]
\newtheorem{proposition}[theorem]{Proposition}
\newtheorem{lemma}[theorem]{Lemma}
\newtheorem{corollary}[theorem]{Corollary}

\theoremstyle{definition}

\newtheorem{example}[theorem]{Example}
\theoremstyle{remark}
\newtheorem{remark}[theorem]{Remark}

\newcommand{\bQ}{\widehat{\mathcal{Q}}}          
\newcommand{\HA}{\mathcal{H}_A}
\newcommand{\HB}{\mathcal{H}_B}
\newcommand{\HR}{\mathcal{H}_R}
\newcommand{\HAp}{\mathcal{H}_{A'}}
\newcommand{\cB}{\mathcal{B}}                     
\newcommand{\base}{\mathfrak{B}}                  
\newcommand{\BHS}{B_{\mathrm{HS}}}                
\newcommand{\cE}{\mathcal{E}}
\newcommand{\cJ}{\mathcal{J}}                     
\newcommand{\cS}{\mathcal{S}}                     
\newcommand{\one}{\mathbf{1}}
\DeclareMathOperator{\cone}{cone}
\DeclareMathOperator{\conv}{conv}
\DeclareMathOperator{\id}{id}
\DeclareMathOperator{\dist}{dist}
\newcommand{\SEP}{\mathrm{SEP}}
\newcommand{\SEPp}{\mathrm{SEP}_{+}}
\newcommand{\opnorm}[1]{\lVert #1 \rVert_{\mathrm{op}}}
\newcommand{\hsnorm}[1]{\lVert #1 \rVert_{\mathrm{HS}}}
\newcommand{\trnorm}[1]{\lVert #1 \rVert_{1}}
\renewcommand{\ip}[2]{\langle #1, #2\rangle}
\newcommand{\WOPT}{\mathrm{WOPT}}
\newcommand{\WMEM}{\mathrm{WMEM}}
\newcommand{\TA}{{T_A}}

\title{\fontsize{15}{18}\selectfont
  \makebox[\textwidth][c]{\textbf{Membership in the cone of quantum states over time is NP-hard}}}
\date{}

\author[1,2]{Matt Hoogsteder-Riera\thanks{\href{mailto:hoogstedermatt@gmail.com}{hoogstedermatt@gmail.com}}}

\affil[1]{\textit{Centre for Quantum Technologies,\protect\\[-1mm] National University of Singapore,\protect\\[-1mm] 3 Science Drive 2, Singapore 117543}}
\affil[2]{\textit{Campus for Research Excellence and Technological Enterprise (CREATE),\protect\\[-1mm] National University of Singapore,\protect\\[-1mm] 1 Create Way, Singapore 13860}}

\begin{document}

\maketitle

\vspace*{-1.2cm}

\begin{abstract}
A state over time (stote) pairs a state $\rho$ with a channel $\cE$ as the Hermitian, generally non-positive operator $(\rho\otimes\one)\star J_\cE$. Characterizing the convex cone $\bQ$ these operators generate was posed as an open problem in~\cite{HoogstederCalsamigliaWinter2025}. We exhibit a linear copy of the bipartite separable cone as an exposed face of $\bQ$: for output systems of the form $B = A'R$ with $A'$ a copy of the input, there is an explicit witness $W_0\in\bQ^*$ and an explicit injective linear map $L$ with $\sigma\in\SEPp(A{:}R)\iff L(\sigma)\in\bQ(A{:}A'R)$. Composing with the NP-hardness of separability shows that exact membership, weak membership, and weak optimization are NP-hard for $\bQ$ and its dual. A quantitative stability bound makes the weak-membership reduction valid at inverse-polynomial accuracy. Thus polynomial-time membership tests do not exist unless $\mathrm{P}=\mathrm{NP}$. Independently of this complexity assumption, Fawzi's theorem implies that $\bQ(A{:}A'R)$ admits no finite semidefinite representation when $d=\dim A\ge2$, $r=\dim R\ge2$, and $d+r>5$.
\end{abstract}

\section{Introduction}\label{sec:intro}

\subsection{States over time and the membership problem}\label{sec:motivation}

For finite classical systems, a joint distribution at two times can be written as $\pi(x,y)=p(x)T(y|x)$, where $p$ is the initial distribution and $T$ is a stochastic map. With both marginals fixed, these distributions form a polytope defined by entrywise nonnegativity and linear marginal constraints. Optimizing a linear cost over this set is the classical optimal transport problem \cite{Villani2009}.

For quantum systems, operators representing temporal correlations need not be positive, as in the pseudo-density-matrix construction of Fitzsimons, Jones and Vedral \cite{FitzsimonsJonesVedral2015}. Horsman, Heunen, Pusey, Barrett and Spekkens \cite{HorsmanHeunenPuseyBarrettSpekkens2017} proved that no construction satisfies all five of their requirements: Hermiticity, compatibility with probabilistic mixing, the correct classical limit, the correct single-time marginals, and associative composition over multiple time steps. Their theorem does not assume positivity. Related formulations include the causally neutral quantum Bayesian framework of Leifer and Spekkens \cite{LeiferSpekkens2013}. Here a \emph{state over time} (stote) is a Hermitian, generally non-positive operator on $\HA\otimes\HB$ that combines an initial state with a channel.

We study the Jordan-product prescription
\begin{equation}\label{eq:jordan-stote}
Q \;=\; (\rho\otimes\one_B)\star J_{\cE}, \qquad X\star Y = \tfrac{1}{2}(XY+YX),
\end{equation}
where $J_{\cE}$ is the Jamio{\l}kowski operator \cite{Jamiolkowski1972,Choi1975} of the channel $\cE\colon\cB(\HA)\to\cB(\HB)$. This prescription is related to the pseudo-density-matrix construction \cite{FitzsimonsJonesVedral2015,HorsmanHeunenPuseyBarrettSpekkens2017} and was developed for general finite-dimensional systems by Fullwood and Parzygnat \cite{FullwoodParzygnat2022}. Their construction avoids the earlier no-go theorem by restricting the domain to the relevant state and channel data, rather than requiring a product on arbitrary operator pairs. Lie and Ng \cite{LieNg2024} subsequently characterized the construction by operational axioms. Our object of study is the convex cone generated by these stotes,
\begin{equation}\label{eq:cone-def}
\bQ(\HA{:}\HB) \;=\; \cone\{(\rho\otimes\one)\star J_{\cE}:\ \rho \text{ a state},\ \cE \text{ a channel}\},
\end{equation}
and its dual cone $\bQ^*$. Here $\cone$ denotes the set of finite conic combinations. We distinguish the generators from their convex hull: a convex mixture of stotes need not itself arise from one state--channel pair, as the exact example in Appendix~\ref{app:checks} shows. Our membership results concern the convex cone, not the generating set alone.

Song and Parzygnat \cite{SongParzygnat2025} prove that every bipartite separable density matrix admits a temporal realization by a state--channel pair, and give necessary and sufficient conditions for temporal compatibility of a bipartite density matrix. Their question concerns an individual realization, whereas ours concerns membership in the convex cone generated by all such stotes.

Goswami, Barvar and Bera \cite[Section~VI.4]{GoswamiBarvarBera2026} give semidefinite tests for compatibility of multi-time states with general causal processes. Those processes allow initial system--environment correlations and memory; the tests do not restrict the process to a convex mixture of the state--channel constructions defining $\bQ$. Their SDP characterization therefore concerns a different set and does not contradict the hardness or non-representability results below.

States over time are used in quantum Bayesian inference and retrodiction \cite{ParzygnatFullwood2023,ParzygnatBuscemi2023}, and related temporal representations occur in causal inference and the reconstruction of dynamics \cite{RiedEtAl2015,Fullwood2025}. Kirkwood--Dirac quasiprobabilities \cite{Kirkwood1933,Dirac1945} provide another description of correlations between incompatible observables; their applications are reviewed in \cite{LostaglioEtAl2023,ArvidssonShukurEtAl2024}. Our direct motivation is quantum optimal transport. Hoogsteder-Riera, Calsamiglia and Winter \cite{HoogstederCalsamigliaWinter2025} define transport costs through Hilbert--Schmidt inner products $\ip{Q}{K}=\Tr[QK]$ and obtain a partial characterization of $\bQ^*$ in their Theorem~4.12. A cost is nonnegative on every stote exactly when its operator belongs to $\bQ^*$. This leads to the characterization problem for $\bQ$ and its dual. Other quantum transport constructions include the semiclassical approach of \cite{GolseMouhotPaul2016} and the channel-based approach of \cite{DePalmaTrevisan2021}.

We prove that exact and weak membership in $\bQ$ and its dual are NP-hard. Thus polynomial-time membership algorithms do not exist in general unless $\mathrm{P}=\mathrm{NP}$. This also excludes uniformly constructible polynomial-size rational SDP formulations satisfying the encoding and conditioning assumptions stated in the proof of Theorem~\ref{thm:weak-hardness}; polynomial size alone is not sufficient for that computational conclusion. Separately, Theorem~\ref{thm:no-lift} rules out every finite semidefinite representation of $\bQ(A{:}A'R)$ when $d,r\ge2$ and $d+r>5$, using Fawzi's theorem \cite{Fawzi2021}.

\subsection{Main results and proof outline}\label{sec:summary}

Our main results establish computational and semidefinite-representation obstructions for the cone of states over time. Fix $\dim\HA = d$. The output system throughout the reduction is a composite $\HB = \HAp\otimes\HR$, with $\HAp$ a copy of $\HA$ and $\dim\HR = r$; write $N := d^2r$ for the total dimension.
\begin{itemize}
\item \emph{Exact membership} in $\bQ$ is NP-hard (Theorem~\ref{thm:exact-hardness}).
\item \emph{Weak membership} in $\bQ$ is strongly NP-hard, with inverse-polynomial accuracy parameters (Theorem~\ref{thm:weak-hardness}). The standard oracle equivalences also give NP-hardness of weak optimization and of weak membership and weak optimization for the dual cone (Corollary~\ref{cor:optimization-dual}; \cite[Ch.~4]{GrotschelLovaszSchrijver1988}). These complexity statements concern variable dimensions.
\item \emph{No finite SDP lift}: $\bQ(A{:}A'R)$ has no finite semidefinite representation when $d,r\ge2$ and $d+r>5$ (Theorem~\ref{thm:no-lift}). This result applies in each such fixed dimension and does not assume $\mathrm{P}\ne\mathrm{NP}$.
\end{itemize}

The main construction identifies a linear copy of the separable cone as an exposed face of $\bQ$. This gives an exact membership reduction, but weak membership requires two further steps: proving that the image of a NO instance of the separability promise problem remains quantitatively separated from the whole target cone, and moving the images of separable inputs from its boundary into its interior. A stability bound for the face gives the first step; mixing with the maximally mixed state gives the second.

The \emph{facial embedding} (Section~\ref{sec:embedding}) uses the witness obtained by tensoring the cost operator $d\one_{AA'}-S_{AA'}$ of \cite[Proposition~5.1]{HoogstederCalsamigliaWinter2025} with $\one_R$,
\begin{equation}\label{eq:witness-preview}
W_0 = \bigl(d\,\one_{AA'} - S_{AA'}\bigr)\otimes\one_R \;\in\; \bQ(A{:}A'R)^{*},
\qquad S_{AA'} = \text{the swap},
\end{equation}
which has zero Hilbert--Schmidt inner product with a generator $(\rho\otimes\one)\star J_\cE$ \emph{exactly} when the channel $\cE$ acts as the identity $A\to A'$ tensored with the preparation of an arbitrary state $\tau$ on $R$ (Lemma~\ref{lem:equality-case}). The corresponding exposed face of $\bQ$ is therefore an isomorphic copy of the separable cone: writing $\Gamma(X) := \tfrac12\{X\otimes\one_{A'}, S_{AA'}\}$ and $L := \Gamma\otimes\id_R$,
\begin{equation}\label{eq:face-preview}
\bQ(A{:}A'R)\,\cap\, W_0^{\perp} \;=\; L\bigl(\SEPp(A{:}R)\bigr),
\end{equation}
and since $L(\sigma)\perp W_0$ holds automatically for \emph{every} Hermitian $\sigma$, membership transfers exactly (Theorem~\ref{thm:facial-embedding}):
\begin{equation}\label{eq:reduction-preview}
\sigma\in\SEPp(A{:}R)
\quad\Longleftrightarrow\quad
L(\sigma)\in\bQ(A{:}A'R).
\end{equation}
The map $L$ is injective, rational, and computable in polynomial time; the target dimensions $d_A = d$, $d_B = dr$ are polynomial in those of the separability instance. Composing with the NP-hardness of separability (Gurvits \cite{Gurvits2003}; in the robust promise form, Gharibian \cite{Gharibian2010}) gives exact-membership hardness. The same facial embedding also transfers Fawzi's obstruction to finite semidefinite representations of $\SEP$ \cite{Fawzi2021}: exposed faces and linear images preserve semidefinite representability, so a representation of $\bQ$ would give one for $\SEP$.

For weak membership, Proposition~\ref{prop:stability} supplies the quantitative estimate: for every $Y\in\bQ$, the trace-norm distance from $Y$ to the face is at most $2\sqrt2\,d^{3/2}\sqrt{\Tr[Y]\,\Tr[W_0Y]}$. This estimate controls how close a point of the cone with a small witness value must be to the embedded separable cone.

\begin{figure}[!htbp]
\centering
\includegraphics[width=\linewidth]{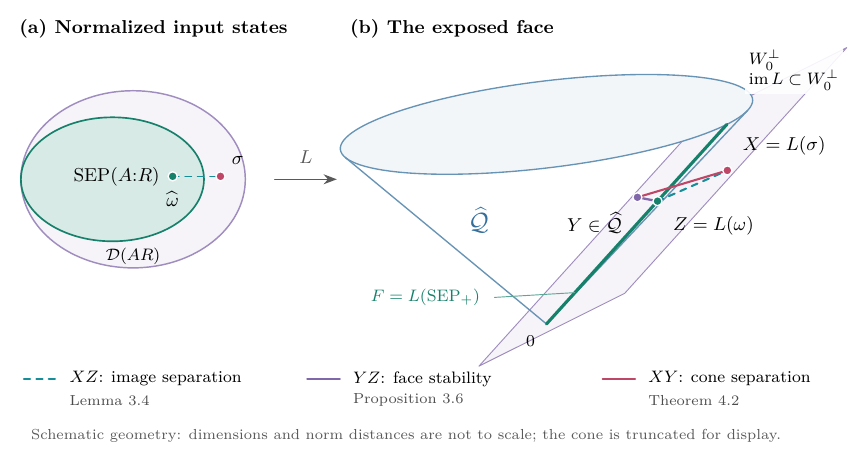}
\caption{Schematic of the facial embedding. The supporting hyperplane $W_0^\perp$ meets $\bQ$ in $F=L(\SEPp)$; $\operatorname{im}L$ is a proper subspace of $W_0^\perp$. The points $X=L(\sigma)$, $Y\in\bQ$, and $Z=L(\omega)\in F$ show the comparisons used in the separation argument. Here $\mathcal D(AR)$ denotes the density operators on $AR$, and $\widehat\omega=\omega/\Tr[\omega]$. The rounded input sets and the ray representing $F$ are schematic, not literal sections of these sets. The right panel suppresses the trace-one constraint; dimensions and distances are not to scale.}
\label{fig:facial-embedding}
\end{figure}

\begin{figure}[!htbp]
\centering
\includegraphics[width=\linewidth]{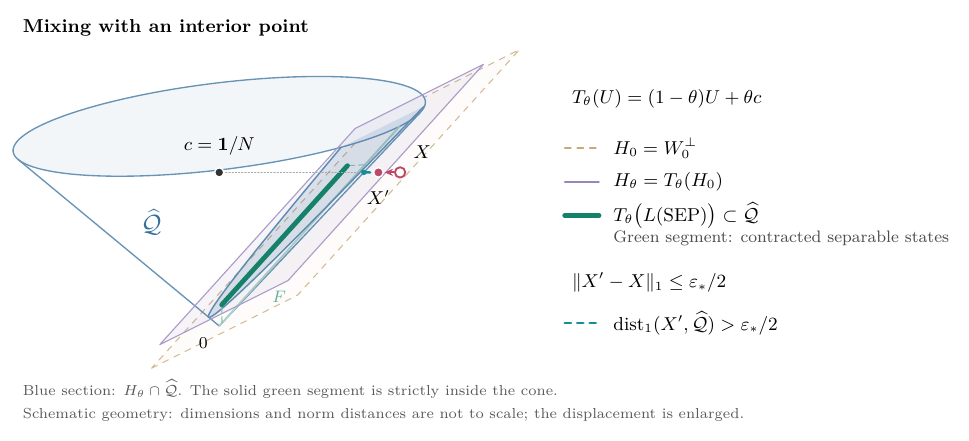}
\caption{Schematic of the mixing step. The map $T_\theta(U)=(1-\theta)U+\theta\one/N$ moves $L(\SEP)$ into the interior of $\bQ$, while the image of a NO instance remains outside. The green segment represents $T_\theta(L(\SEP))$; the blue section is $T_\theta(W_0^\perp)\cap\bQ$. The displacement is enlarged for display.}
\label{fig:mixing-step}
\end{figure}

\paragraph{Main ideas of the proof.} Figures~\ref{fig:facial-embedding} and~\ref{fig:mixing-step} illustrate the reduction in Theorem~\ref{thm:weak-hardness}. The map $L$ sends Hermitian operators injectively into the supporting hyperplane $W_0^\perp$, whose intersection with $\bQ$ is exactly $L(\SEPp)$. Its image is a proper subspace of that hyperplane, not the whole hyperplane. A separable input state therefore maps to the boundary of the trace-one base $\base$. To obtain the interior margin required by weak membership, we apply $T_{\theta}(X)=(1-\theta)X+\theta\one/N$. Convexity and a separable ball around $\one/N$ ensure that $T_\theta(L(\SEP))$ lies in the relative interior of $\base$. The mixing parameter is small enough to preserve an inverse-polynomial separation for NO instances. The construction is polynomial-time, so a weak-membership oracle for $\bQ$ decides the source separability promise problem.

The main technical step is the NO case. If $L(\sigma)$ were close to some $Y\in\bQ$, the identity $\Tr[W_0L(\sigma)]=0$ would make $\Tr[W_0Y]$ small. The stability bound would then place $Y$ close to $L(\SEPp)$. Taking the partial trace over $A'$ and normalizing would give a separable state too close to $\sigma$, contradicting the source promise. This argument proves separation from the whole cone, not only from the embedded separable set. The final mixing step reduces that separation by at most a factor of two.

\subsection{Definitions, notation, and basic identities}\label{sec:conventions}

We follow \cite{HoogstederCalsamigliaWinter2025}. All spaces are finite-dimensional; $d_A = \dim\HA$, $d_B = \dim\HB$, $N = d_Ad_B$; in the reduction sections, $\HB = \HAp\otimes\HR$ with $\dim\HAp = \dim\HA = d$ and $\dim\HR = r$, so $N = d^2r$.

We write $\operatorname{Herm}(\mathcal H)$ for the real vector space of Hermitian operators on $\mathcal H$, $\cS(\mathcal H)$ for its positive semidefinite operators of trace one, and $P_\psi:=\ketbra{\psi}{\psi}$ for the pure state associated with a unit vector $\psi$. The symbols $\one_A$ and $\id_A$ denote the identity operator on $\HA$ and the identity map on $\cB(\HA)$, respectively. We identify $A$ and its copy $A'$ through their computational bases and write $\cE\colon A\to B$ as shorthand for $\cE\colon\cB(\HA)\to\cB(\HB)$. The anticommutator is $\{X,Y\}:=XY+YX$.

The Choi matrix \cite{Choi1975} of a CP map $\cE\colon\cB(\HA)\to\cB(\HB)$ is
\begin{equation}\label{eq:choi-def}
C_\cE = \sum_{ij}\ketbra{i}{j}\otimes\cE(\ketbra{i}{j}) \;\succeq\; 0,
\end{equation}
with trace preservation $\iff \Tr_B[C_\cE] = \one_A$; the Jamio{\l}kowski matrix \cite{Jamiolkowski1972} is $J_\cE = C_\cE^{\TA}$ ($\TA$ = partial transpose on $A$ in the computational basis), equivalently
\begin{equation}\label{eq:jam-explicit}
J_\cE = \sum_{ij}\ketbra{i}{j}\otimes\cE(\ketbra{j}{i}),
\end{equation}
and $\cJ = \cJ(\HA\to\HB)$ denotes the (convex, spectrahedral) set of Jamio{\l}kowski matrices of channels, $\cJ = \{\tilde C^{\TA} : \tilde C\succeq 0,\ \Tr_B[\tilde C] = \one_A\}$. The stote of $(\rho,\cE)$ is $Q_{\rho,\cE} = (\rho\otimes\one_B)\star J_\cE$; it satisfies $\Tr_B[Q] = \rho$, $\Tr_A[Q] = \cE(\rho)$, $\Tr[Q] = 1$. We write
\begin{equation}\label{eq:cone-base-def}
\bQ = \cone\{(\rho\otimes\one)\star J:\ \rho\in\cS(\HA),\ J\in\cJ\},
\qquad
\base = \conv\{(\rho\otimes\one)\star J\}\cap\{\Tr = 1\},
\end{equation}
for the cone and its base. Operators are compared in the operator norm $\opnorm{\cdot}$, the trace norm $\trnorm{\cdot}$, and the Hilbert--Schmidt norm $\hsnorm{X}=\sqrt{\Tr[X^\dagger X]}$, induced by $\ip{X}{Y}=\Tr[X^\dagger Y]$. The unsubscripted norm $\norm{v}$ denotes the Hilbert-space norm of a vector. We write $\dist_1(X,C):=\inf_{Y\in C}\trnorm{X-Y}$ and $\dist_{\mathrm{HS}}(X,C):=\inf_{Y\in C}\hsnorm{X-Y}$ for distances to a set. The notation $\BHS(x,s)$ denotes the closed Hilbert--Schmidt ball of radius $s$ about $x$; $\cB(\mathcal H)$ denotes the operator algebra and $\base$ the trace-one base of $\bQ$. We write $\SEP(A{:}B)$ for the trace-one separable states across $A{:}B$ and $\SEPp(A{:}B)=\cone\SEP(A{:}B)$ for the closed separable cone, and similarly for $A{:}R$. When the bipartition is omitted, it is the one under discussion.

We also recall the definition of an exposed face \cite[\S 18]{Rockafellar1970}. Let $C$ be a convex set in a finite-dimensional real inner-product space. A nonempty subset $F\subseteq C$ is an \emph{exposed face} if there are a real linear functional $\ell$ and a scalar $\alpha$ such that $\ell(x)\ge\alpha$ for every $x\in C$ and $F=\{x\in C:\ell(x)=\alpha\}$. Thus $F$ is the part of $C$ on which a supporting functional attains its minimum. For a convex cone $C$, its dual cone is $C^*:=\{w:\ip{x}{w}\ge0\text{ for every }x\in C\}$. If $w\in C^*$, then
\begin{equation}\label{eq:exposed-cone-face}
F_w=C\cap w^\perp
=\{x\in C:\ip{x}{w}=0\}
=\operatorname*{arg\,min}_{x\in C}\ip{x}{w}
\end{equation}
is an exposed face; it is nonempty because $0\in C$. This notion is central to the main argument. Section~\ref{sec:embedding} constructs $W_0\in\bQ^*$ and proves that the exposed face $F_{W_0}$ is a linear copy of the separable cone. This identification gives the membership reductions. Section~\ref{sec:consequences} also uses the fact that exposed faces and linear images preserve semidefinite representability to obtain the non-representability result.

Throughout, $S=S_{AA'}=\sum_{ij}\ketbra{ij}{ji}$ swaps $A$ and $A'$, and $\ket{\Omega}=\sum_i\ket{ii}$ is the unnormalized maximally entangled vector ($\norm{\Omega}^2=d$). From \eqref{eq:jam-explicit}, the identity channel $A\to A'$ has $C_{\id}=\ketbra{\Omega}{\Omega}$ and $J_{\id}=S$.

The following four swap and vectorization identities will be used below; they follow directly from $S=\sum_{ij}\ketbra{ij}{ji}$ and $\ket{\Omega}=\sum_i\ket{ii}$ (see also \cite[\S 1.1.2]{Watrous2018}). For $X,Z$ on $\mathbb C^d$, $Y$ on $\HAp\otimes\HR$, and $M\colon\HA\to\mathcal H$ arbitrary,
\begin{align}
&\Tr[S(X\otimes Z)] = \Tr[XZ], \label{eq:swap-trace}\\
&\Tr_{A'}\bigl[(X\otimes\one_{A'})S\bigr]
 = \Tr_{A'}\bigl[S(X\otimes\one_{A'})\bigr] = X, \label{eq:swap-partial}\\
&\Tr\bigl[(S\otimes\one_R)(X_A\otimes Y_{A'R})\bigr]
 = \Tr\bigl[(X_{A'}\otimes\one_R)\,Y_{A'R}\bigr], \label{eq:swap-three}\\
&\bigl\lVert(\one_A\otimes M)\ket{\Omega}\bigr\rVert^2 = \Tr[M^\dagger M].
\label{eq:omega-isometry}
\end{align}

\section{Cone geometry and weak computational problems}\label{sec:geometry}

This section establishes the geometric facts needed by the reduction. Product states generate the separable cone inside $\bQ$, and the Gurvits--Barnum theorem gives an explicit ball in its trace-one base (Lemma~\ref{lem:prepare-generators}). The cone is closed, pointed, and spanning, and every element is a finite conic combination of generators (Proposition~\ref{prop:cone-properties}). The separable set over the same bipartition cannot lie in a proper exposed face (Remark~\ref{rem:no-exposing-witness}). Finally, inner and outer ball bounds justify the weak-oracle reductions used in Section~\ref{sec:consequences} (Proposition~\ref{prop:balls}). Appendix~\ref{app:checks} gives an exact example showing that the generating set is not convex.

\begin{lemma}[product states are stotes; the separable ball]\label{lem:prepare-generators}
For every unit vector $\psi\in\HA$ and state $\sigma$ on $\HB$, the product state $P_\psi\otimes\sigma$ is a stote generator: it is the stote of the input $P_\psi$ and the prepare channel $\cE_\sigma(X) = \Tr[X]\,\sigma$. Consequently
\begin{equation}\label{eq:sep-in-cone}
\SEPp(A{:}B) \;\subseteq\; \bQ(A{:}B),
\end{equation}
and the Gurvits--Barnum ball \cite{GurvitsBarnum2002} lies inside the base:
\begin{equation}\label{eq:gb-ball}
\BHS\Bigl(\tfrac{\one}{N},\ \tfrac{1}{\sqrt{N(N-1)}}\Bigr)\cap\{\Tr = 1\}
\;\subseteq\; \SEP\cap\{\Tr = 1\} \;\subseteq\; \base .
\end{equation}
\end{lemma}

\begin{proof}
For the prepare channel $\cE_\sigma(X) = \Tr[X]\,\sigma$, the Choi matrix is $\tilde C = \one_A\otimes\sigma$ ($\succeq 0$, $\Tr_B[\tilde C] = \one_A$, so $\cE_\sigma$ is a channel), and its Jamio{\l}kowski matrix $J = \tilde C^{\TA} = \one_A\otimes\sigma$ is unchanged. For a pure input $P_\psi$,
\begin{equation}\label{eq:prepare-product}
(P_\psi\otimes\one_B)\star(\one_A\otimes\sigma)
= \tfrac12\bigl((P_\psi\otimes\one)(\one\otimes\sigma) + (\one\otimes\sigma)(P_\psi\otimes\one)\bigr)
= P_\psi\otimes\sigma,
\end{equation}
since the two factors act on different tensor slots and hence commute. So every pure product state $P_\psi\otimes P_\phi$ is a generator; taking conic and convex combinations gives \eqref{eq:sep-in-cone} and $\SEP\cap\{\Tr=1\}\subseteq\base$. The ball statement is then the Gurvits--Barnum theorem \cite{GurvitsBarnum2002}: every Hermitian $\omega$ with $\Tr[\omega] = 1$ and $\hsnorm{\omega - \one/N}\le 1/\sqrt{N(N-1)}$ is separable; chain the inclusions.
\end{proof}

\begin{proposition}[cone properties]\label{prop:cone-properties}
The following hold:
\begin{itemize}
\item[(i)] every generator $(\rho\otimes\one)\star J$ has unit trace, and $\base$ is compact and convex with $0\notin\base$ --- a base of $\bQ$ in the sense of \cite[Ch.~1]{AliprantisTourky2007};
\item[(ii)] $\bQ$ is closed, pointed and spanning, and every element of $\bQ$ is a \emph{finite} conic combination of generators (at most $N^2 + 1$ of them, by Carath\'eodory);
\item[(iii)] $\bQ$ is invariant under $Q\mapsto(U\otimes V)Q(U\otimes V)^\dagger$ for all unitaries $U, V$.
\end{itemize}
\end{proposition}

\begin{proof}
(i) \emph{Unit trace.} By cyclicity $\Tr[Q] = \Tr[(\rho\otimes\one)J]$, and by trace duality of the partial transpose ($\Tr[XY^{\TA}] = \Tr[X^{\TA}Y]$), $\Tr[(\rho\otimes\one)\tilde C^{\TA}] = \Tr[(\rho^T\otimes\one)\tilde C] = \Tr[\rho^T\,\Tr_B[\tilde C]] = \Tr[\rho^T] = 1$.

\emph{Compactness.} The parameter set $\cS(\HA)\times\cJ$ is compact: states form a closed bounded set, and $\cJ$ is the image of the closed bounded set $\{\tilde C\succeq 0,\ \Tr_B[\tilde C] = \one_A\}$ (bounded because $\tilde C\succeq 0$ and $\Tr[\tilde C] = d_A$ force $\opnorm{\tilde C}\le d_A$) under the linear homeomorphism $\TA$. The map $(\rho, J)\mapsto(\rho\otimes\one)\star J$ is continuous (bilinear on finite-dimensional spaces), so the generating set $G := \{(\rho\otimes\one)\star J\}$ is compact, and $\base = \conv(G)$ is compact by Carath\'eodory (the convex hull of a compact set in $\mathbb{R}^n$ is compact \cite{Rockafellar1970}). Every element of $\base$ has unit trace, so $0\notin\base$; since $\bQ = \bigcup_{t\ge0}t\,\base$ and every nonzero element has the unique representation $t\,b$ with $t$ equal to the trace of the element, $\base$ is a base.

(ii) \emph{Closedness.} A cone generated by a compact convex set not containing the origin is closed (cf.\ \cite[\S 9]{Rockafellar1970}, \cite[Ch.~1]{AliprantisTourky2007}): if $t_nb_n\to x$ with $b_n\in\base$, $t_n\ge0$, then $t_n = \Tr[t_nb_n]\to\Tr[x] =: t\ge0$; if $t>0$ then $b_n\to x/t\in\base$ by closedness of $\base$ and $x = t(x/t)\in\bQ$, while if $t = 0$ then $\hsnorm{t_nb_n}\le t_n\sup_{b\in\base}\hsnorm{b}\to0$, so $x = 0\in\bQ$. \emph{Pointedness:} $Q, -Q\in\bQ$ with $Q\ne0$ would give $\Tr[Q]>0$ and $\Tr[-Q]>0$, a contradiction. \emph{Spanning:} $G$ contains all pure product states (Lemma~\ref{lem:prepare-generators}), which span the Hermitian operators (rank-one projectors span the Hermitian operators of each factor --- the projectors onto $\ket{i}$, $\ket{j}$, $(\ket{i}+\ket{j})/\sqrt2$, $(\ket{i}+i\ket{j})/\sqrt2$ already span the $(i,j)$ matrix-unit directions --- and tensor products of spanning sets span the tensor product), so $\bQ - \bQ = \operatorname{Herm}(\HA\otimes\HB)$. \emph{Finite combinations:} every $x\in\bQ$, $x\ne0$, is $\Tr[x]\cdot b$ with $b\in\base = \conv(G)$; Carath\'eodory in dimension $N^2$ writes $b$ as a convex combination of at most $N^2+1$ generators.

(iii) Write $W := U\otimes V$ and $\mathrm{Ad}_W(X) := WXW^\dagger$. Inserting $W^\dagger W = \one$ between the two factors of each term of the Jordan product,
\begin{equation}\label{eq:lu-covariance}
\mathrm{Ad}_W\bigl((\rho\otimes\one)\star J\bigr)
= (U\rho U^\dagger\otimes\one)\star\mathrm{Ad}_W(J),
\end{equation}
using $W(\rho\otimes\one)W^\dagger = U\rho U^\dagger\otimes\one$. It remains to check $\mathrm{Ad}_{U\otimes V}(J)\in\cJ$: in Choi coordinates, $\mathrm{Ad}_{U\otimes V}(\tilde C^{\TA}) = \bigl((\bar U\otimes V)\tilde C(\bar U\otimes V)^\dagger\bigr)^{\TA}$ (valid since $\TA\circ\mathrm{Ad}_U = \mathrm{Ad}_{\bar U}\circ\TA$ on the $A$ factor); the new Choi matrix is PSD with $\Tr_B$-marginal $\bar U\one_A\bar U^\dagger = \one_A$. Hence conjugated generators are generators, and invariance of $\bQ$ follows by linearity and closure.
\end{proof}

\begin{remark}[no witness exposes the separable part]\label{rem:no-exposing-witness}
No nonzero $K\in\bQ(A{:}B)^*$ vanishes on $\SEP(A{:}B)$. Hence $\SEP(A{:}B)$ is not contained in a proper exposed face of $\bQ(A{:}B)$: over a \emph{fixed} bipartition, no reduction from separability can proceed by cutting $\SEP$ out of $\bQ$ with a supporting hyperplane.

The set $\SEP$ spans the real vector space of Hermitian operators on $\HA\otimes\HB$, as shown in the proof of Proposition~\ref{prop:cone-properties}(ii). A linear functional vanishing on a spanning set vanishes identically, so $K=0$ by nondegeneracy of the Hilbert--Schmidt pairing. By the definition in Section~\ref{sec:conventions}, a proper exposed face of the cone has the form $F_K=\{Q\in\bQ:\ip{Q}{K}=0\}$ for some nonzero $K\in\bQ^*$, so it cannot contain $\SEP$.
\end{remark}

This observation clarifies the possible form of a facial reduction. It rules out presenting $\SEP(A{:}B)$ itself as facial structure of $\bQ(A{:}B)$; it does not rule out presenting an isomorphic \emph{image} of a separable cone over a \emph{smaller} bipartition as a face, using extra output registers. Section~\ref{sec:embedding} does exactly that: the copy $L(\SEPp(A{:}R))$ sits inside $\bQ(A{:}A'R)$, and the spanning argument above does not apply to it because the image of $L$ is a proper subspace.

\paragraph{Weak problems.}
We use weak optimization and weak membership in the sense of \cite[\S 2.1 and Ch.~4]{GrotschelLovaszSchrijver1988}. In its approximate-value form, $\WOPT(\bQ)$ asks, given a rational Hermitian $K$ and a rational $\varepsilon>0$, for a number within $\varepsilon$ of the support function $h(K)=\max_{\omega\in\base}\ip{\omega}{K}$. This value problem is polynomial-time oracle equivalent to the usual weak-optimization formulations for the well-bounded body considered here. The problem $\WMEM(\bQ)$ asks, given a rational Hermitian $\omega$ of trace one and a rational $\varepsilon>0$, to assert either that $\omega$ lies within distance $\varepsilon$ of $\base$ or that it is not in the $\varepsilon$-interior of $\base$. Here the $\varepsilon$-interior consists of points whose closed $\varepsilon$-ball is contained in $\base$; distances and balls use the Hilbert--Schmidt metric within the trace-one hyperplane. Both answers are allowed when both assertions hold. Sizes are measured in the binary encoding length of the input, including the accuracy parameter; a rational complex entry has rational real and imaginary parts.

\begin{proposition}[GLS well-posedness]\label{prop:balls}
Within $\{\Tr = 1\}$,
\begin{equation}\label{eq:gls-balls}
\BHS\bigl(\one/N,\ 1/\sqrt{N(N-1)}\bigr)\ \subseteq\ \base\ \subseteq\ \BHS(0, d_A).
\end{equation}
Hence $\base$ is a well-bounded centered convex body in the sense of \cite[\S 2.1]{GrotschelLovaszSchrijver1988}, and the oracle reductions between the weak problems for $\base$, its polar, and linear optimization over either --- the shallow-cut ellipsoid machinery of Gr{\"o}tschel--Lov{\'a}sz--Schrijver \cite[Ch.~4]{GrotschelLovaszSchrijver1988}, originating with Yudin--Nemirovski \cite{YudinNemirovski1976} --- all run in oracle-polynomial time with polynomially related accuracy parameters.
\end{proposition}

\begin{proof}
\emph{Inner ball:} Lemma~\ref{lem:prepare-generators}; the radius $1/\sqrt{N(N-1)}\ge1/N$ is inverse-polynomial. \emph{Outer ball:} for any generator, first bound the Jamio{\l}kowski matrix: $\hsnorm{J} = \hsnorm{\tilde C^{\TA}} = \hsnorm{\tilde C}$ (the partial transpose permutes matrix entries, preserving the Hilbert--Schmidt norm), and $\hsnorm{\tilde C}\le\trnorm{\tilde C} = \Tr[\tilde C] = d_A$ for $\tilde C\succeq 0$. Then, writing $P := \rho\otimes\one$ (a positive contraction),
\begin{equation}\label{eq:generator-hs-bound}
\hsnorm{(\rho\otimes\one)\star J}
\le \tfrac12\hsnorm{PJ} + \tfrac12\hsnorm{JP}
\le \opnorm{P}\,\hsnorm{J} \le d_A,
\end{equation}
using $\hsnorm{XY}\le\opnorm{X}\hsnorm{Y}$; convex combinations obey the same bound. \emph{Well-boundedness:} translating by the center $\one/N$, the body $\base - \one/N$ (inside the trace-zero hyperplane, of real dimension $N^2-1$) contains the ball of radius $1/N$ and is contained in the ball of radius $d_A + 1$; the ratio is polynomial in $N$, which is the hypothesis of the oracle framework. The listed reductions are \cite[Thm.~4.3.2 and \S\S 4.4--4.7]{GrotschelLovaszSchrijver1988}.
\end{proof}

\section{Facial embedding and stability}\label{sec:embedding}

Throughout this section the output is the composite $\HB = \HAp\otimes\HR$, with $\HAp\cong\HA\cong\mathbb{C}^d$ and $\HR\cong\mathbb{C}^r$, and
\begin{equation}\label{eq:witness-def}
W_0 := \bigl(d\,\one_{AA'} - S_{AA'}\bigr)\otimes\one_R .
\end{equation}
For a channel $\cE\colon A\to A'R$ we write $\Lambda := \Tr_R\circ\,\cE\colon A\to A'$ for its reduced channel. The section proves, in order: an exact formula for the pairing of $W_0$ with a generator, in which only $\Lambda$ appears (Lemma~\ref{lem:pairing}); nonnegativity, i.e.\ $W_0\in\bQ^*$ (Lemma~\ref{lem:witness}); rigidity of the equality case --- zero pairing forces $\cE = \id\otimes\tau$ (Lemma~\ref{lem:equality-case}); the properties of the embedding map $L$ (Lemma~\ref{lem:embedding-map}); the facial identification and the exact reduction (Theorem~\ref{thm:facial-embedding}); and a quantitative stability bound for the face (Proposition~\ref{prop:stability}), which is what Section~\ref{sec:consequences} uses to make the reduction robust.

\begin{lemma}[pairing formula]\label{lem:pairing}
Let $\rho$ be a state on $\HA$ and $\cE\colon A\to A'R$ a channel. For any Kraus representation $\Lambda(X) = \sum_k K_kXK_k^\dagger$, $\sum_kK_k^\dagger K_k = \one$, of the reduced channel,
\begin{equation}\label{eq:pairing}
\Tr\bigl[W_0\,Q_{\rho,\cE}\bigr]
= d \;-\; \operatorname{Re}\sum_k\overline{\Tr[K_k]}\;\Tr[\rho K_k].
\end{equation}
The value of the right-hand side depends only on the reduced channel $\Lambda$, not on the chosen Kraus representation; see also \cite[Remark~5.3]{HoogstederCalsamigliaWinter2025}.
\end{lemma}

\begin{proof}
\emph{Identity part.} Stotes have unit trace (Proposition~\ref{prop:cone-properties}(i)), so the $d\,\one$ part of $W_0$ contributes exactly $d$.

\emph{Swap part.} By cyclicity,
\begin{equation}\label{eq:swap-split}
\Tr\bigl[(S\otimes\one_R)\,\tfrac12\{\rho\otimes\one, J_\cE\}\bigr]
= \tfrac12\Tr\bigl[(S\otimes\one_R)(\rho\otimes\one)J_\cE\bigr]
+ \tfrac12\Tr\bigl[(S\otimes\one_R)J_\cE(\rho\otimes\one)\bigr],
\end{equation}
and the two terms are complex conjugates of one another (all three operators are Hermitian: the adjoint of $S(\rho\otimes\one)J$ is $J(\rho\otimes\one)S$, whose trace is the second term by cyclicity), so their sum is the real part of twice the first. For arbitrary Hermitian $M$ on $\HA$, inserting \eqref{eq:jam-explicit} and applying \eqref{eq:swap-three} term by term,
\begin{equation}\label{eq:swap-reduces}
\Tr\bigl[(S\otimes\one_R)(M\otimes\one)J_\cE\bigr]
= \sum_{ij}\Tr\bigl[(M\ketbra{i}{j}\otimes\one_R)\,\cE(\ketbra{j}{i})\bigr]
= \sum_{ij}\Tr\bigl[M\ketbra{i}{j}\,\Lambda(\ketbra{j}{i})\bigr],
\end{equation}
because the operator in the first slot acts trivially on $R$; hence the expression depends only on $\Tr_R\circ\cE = \Lambda$. Using a Kraus representation of $\Lambda$, the last expression in \eqref{eq:swap-reduces} becomes
\begin{equation}\label{eq:kraus-reduces}
\begin{aligned}
\sum_{ij}\Tr\bigl[M\ketbra{i}{j}\,\Lambda(\ketbra{j}{i})\bigr]
&= \sum_{ijk}\Tr\bigl[M\ketbra{i}{j}K_k\ketbra{j}{i}K_k^\dagger\bigr]\\
&= \sum_{ijk}\bra{j}K_k\ket{j}\,\bra{i}K_k^\dagger M\ket{i}\\
&= \sum_k\Tr[K_k]\;\Tr[K_k^\dagger M].
\end{aligned}
\end{equation}
Setting $M = \rho$ and using $\Tr[K^\dagger\rho] = \overline{\Tr[\rho K]}$,
\begin{equation}\label{eq:swap-value}
\Tr\bigl[(S\otimes\one_R)\,Q_{\rho,\cE}\bigr]
= \operatorname{Re}\sum_k\Tr[K_k]\;\overline{\Tr[\rho K_k]}
= \operatorname{Re}\sum_k\overline{\Tr[K_k]}\;\Tr[\rho K_k].
\end{equation}
Kraus-independence follows from the middle expression of \eqref{eq:swap-reduces}, which is defined by $\Lambda$ alone.
\end{proof}

\begin{lemma}[$W_0$ is a witness]\label{lem:witness}
$\Tr[W_0\,Q_{\rho,\cE}]\ge0$ for every state $\rho$ and channel $\cE$; hence $W_0\in\bQ(A{:}A'R)^*$.
\end{lemma}

\begin{proof}
By Lemma~\ref{lem:pairing}, it is enough to prove $\operatorname{Re}\sum_k\overline{a_k}b_k\le d$, where $a_k := \Tr[K_k]$ and $b_k := \Tr[\rho K_k]$. Cauchy--Schwarz against $\one$ gives $\abs{a_k}^2\le d\,\Tr[K_k^\dagger K_k]$, so
\begin{equation}\label{eq:a-bound}
\sum_k\abs{a_k}^2 \;\le\; d\sum_k\Tr[K_k^\dagger K_k] \;=\; d\,\Tr[\one] \;=\; d^2 .
\end{equation}
Applying Hilbert--Schmidt Cauchy--Schwarz to the pair $(\sqrt\rho,\ K_k\sqrt\rho)$ gives
\begin{equation}\label{eq:b-term-bound}
\abs{b_k}^2\le\Tr[\rho]\,\Tr[\rho K_k^\dagger K_k]
= \Tr[\rho K_k^\dagger K_k].
\end{equation}
Summing and using trace preservation,
\begin{equation}\label{eq:b-bound}
\sum_k\abs{b_k}^2 \;\le\; \Tr\Bigl[\rho\sum_kK_k^\dagger K_k\Bigr] \;=\; \Tr[\rho] \;=\; 1 .
\end{equation}
By Cauchy--Schwarz on the $k$-sum, $\abs{\sum_k\bar a_kb_k}\le\sqrt{d^2}\cdot\sqrt1 = d$, so Lemma~\ref{lem:pairing} gives $\Tr[W_0Q_{\rho,\cE}]\ge d-d=0$. Nonnegativity extends to conic combinations and to limits, hence to all of $\bQ$.
\end{proof}

\begin{lemma}[equality case]\label{lem:equality-case}
$\Tr[W_0\,Q_{\rho,\cE}] = 0$ if and only if $\cE(X) = X\otimes\tau$ for some state $\tau$ on $\HR$, where $X$ is identified with the corresponding operator on $A'$. In that case $J_\cE = S\otimes\tau$ and
\begin{equation}\label{eq:face-generator}
Q_{\rho,\cE} = \Gamma(\rho)\otimes\tau,
\qquad
\Gamma(X) := \tfrac12\bigl\{X\otimes\one_{A'},\ S\bigr\}.
\end{equation}
\end{lemma}

\begin{proof}
(\emph{If.}) For $\cE(X) = X\otimes\tau$ the reduced channel is the identity, with single Kraus operator $K = \one$, so \eqref{eq:pairing} gives $d - \operatorname{Re}\bigl[\overline{\Tr[\one]}\,\Tr[\rho\,\one]\bigr] = d - d = 0$. Moreover $C_\cE = \ketbra{\Omega}{\Omega}\otimes\tau$, so $J_\cE = (\ketbra{\Omega}{\Omega})^{\TA}\otimes\tau = S\otimes\tau$, and since the anticommutator acts on $AA'$ while $\tau$ multiplies on $R$, $Q_{\rho,\cE} = \tfrac12\{\rho\otimes\one_{A'R},\ S\otimes\tau\} = \Gamma(\rho)\otimes\tau$.

(\emph{Only if.}) Zero pairing forces $\operatorname{Re}\sum_k\bar a_kb_k = d$, so equality holds throughout the inequalities in the proof of Lemma~\ref{lem:witness}. In particular, $\sqrt{\sum\abs{a_k}^2}\cdot\sqrt{\sum\abs{b_k}^2} = d$ with the factors bounded by $d$ and $1$ respectively, forcing $\sum_k\abs{a_k}^2 = d^2$ (and $\sum_k\abs{b_k}^2 = 1$). Now
\begin{equation}\label{eq:termwise}
\sum_k\Bigl(d\,\Tr[K_k^\dagger K_k] - \abs{\Tr[K_k]}^2\Bigr) \;=\; d^2 - d^2 \;=\; 0
\end{equation}
is a sum of nonnegative terms (Cauchy--Schwarz against $\one$), so each vanishes; equality in Cauchy--Schwarz forces $K_k = c_k\one$ for every $k$, and trace preservation gives $\sum_k\abs{c_k}^2 = 1$. Hence $\Lambda = \id$.

To lift this to $\cE$: $\Tr_R[C_\cE] = C_\Lambda = C_{\id} = \ketbra{\Omega}{\Omega}$, a rank-one operator. For any $\ket{v}\in\mathcal{H}_{AA'}$ orthogonal to $\ket{\Omega}$, the compression $(\bra{v}\otimes\one_R)\,C_\cE\,(\ket{v}\otimes\one_R)\succeq0$ has trace $\bra{v}\Tr_R[C_\cE]\ket{v} = \abs{\braket{v}{\Omega}}^2 = 0$, hence vanishes; positivity then gives $C_\cE(\ket{v}\otimes\HR) = 0$. So $C_\cE$ is supported on $\mathbb{C}\ket{\Omega}\otimes\HR$: $C_\cE = \ketbra{\Omega}{\Omega}\otimes\tau$ with $\tau\succeq0$, and $\Tr[C_\cE] = d$ forces $\Tr[\tau] = 1$. Reading off the channel from the Choi blocks, $\cE(\ketbra{i}{j}) = (\bra{i}\otimes\one)C_\cE(\ket{j}\otimes\one) = \ketbra{i}{j}_{A'}\otimes\tau$, i.e.\ $\cE(X) = X\otimes\tau$.
\end{proof}

\begin{lemma}[the embedding map]\label{lem:embedding-map}
Let $\operatorname{Herm}(\mathcal H)$ denote the real vector space of Hermitian operators on $\mathcal H$. Define the real-linear map
\begin{equation}\label{eq:L-domain-codomain}
L := \Gamma\otimes\id_R\colon
\operatorname{Herm}(\HA\otimes\HR)
\longrightarrow
\operatorname{Herm}(\HA\otimes\HAp\otimes\HR),
\end{equation}
with $\Gamma$ as in \eqref{eq:face-generator}. Then:
\begin{itemize}
\item[(i)] $\Tr_{A'}\circ\,\Gamma = \id$; hence $(\Tr_{A'}\otimes\id_R)\circ L = \id$, $L$ is injective, and $\Tr[L(\sigma)] = \Tr[\sigma]$;
\item[(ii)] $\Tr[W_0\,L(\sigma)] = 0$ for \emph{every} Hermitian $\sigma$;
\item[(iii)] $\trnorm{\sigma}\le\trnorm{L(\sigma)}\le d\,\trnorm{\sigma}$;
\item[(iv)] in the computational basis $L$ has rational entries and is computable in time polynomial in $d, r$ and the bit size of the input.
\end{itemize}
\end{lemma}

\begin{proof}
(i) For every Hermitian $X$ on $\HA$, the definition of $\Gamma$ and \eqref{eq:swap-partial} give
\begin{equation}\label{eq:Gamma-left-inverse}
\begin{aligned}
\Tr_{A'}[\Gamma(X)]
&= \tfrac12\Tr_{A'}\bigl[(X\otimes\one_{A'})S\bigr]
 + \tfrac12\Tr_{A'}\bigl[S(X\otimes\one_{A'})\bigr] \\
&= \tfrac12 X + \tfrac12 X = X.
\end{aligned}
\end{equation}
Thus $\Tr_{A'}\circ\Gamma=\id$. Tensoring this identity with $\id_R$ yields, for every Hermitian $\sigma$ on $\HA\otimes\HR$,
\begin{equation}\label{eq:L-left-inverse}
(\Tr_{A'}\otimes\id_R)[L(\sigma)]
= \bigl((\Tr_{A'}\circ\Gamma)\otimes\id_R\bigr)(\sigma)
= (\id_A\otimes\id_R)(\sigma)=\sigma.
\end{equation}
Hence $L(\sigma)=0$ implies $\sigma=0$, so $L$ is injective. Taking the trace in \eqref{eq:L-left-inverse} also gives $\Tr[L(\sigma)]=\Tr[\sigma]$.

(ii) By linearity it suffices to take $\sigma = M\otimes T$, with $M$ Hermitian on $A$ and $T$ Hermitian on $R$; then $\Tr[W_0L(\sigma)] = \Tr[(d\one - S)\Gamma(M)]\cdot\Tr[T]$. By \eqref{eq:swap-partial}, $\Tr[\Gamma(M)] = \Tr[M]$. Using $S^2 = \one$ and cyclicity one obtains
\begin{equation}\label{eq:S-Gamma-trace}
\Tr[S\,\Gamma(M)]
= \tfrac12\Tr[S(M\otimes\one)S] + \tfrac12\Tr[S^2(M\otimes\one)]
= \tfrac12\,d\Tr[M] + \tfrac12\,d\Tr[M] = d\,\Tr[M],
\end{equation}
so $\Tr[(d\one - S)\Gamma(M)] = d\Tr[M] - d\Tr[M] = 0$. 

(iii) The lower bound follows from trace-norm contractivity of the partial trace on Hermitian operators \cite[Cor.~3.40]{Watrous2018} together with (i). For the upper bound, place the tensor factors in the order $AA'R$ and regard $\sigma_{AR}$ as acting trivially on $A'$. Then
\begin{equation}\label{eq:l-anticommutator}
L(\sigma)=\tfrac12\bigl\{\sigma_{AR}\otimes\one_{A'},\ S_{AA'}\otimes\one_R\bigr\},
\end{equation}
where the tensor placement is implicit. The trace-norm ideal property \cite[Eq.~(1.175)]{Watrous2018}, together with the triangle inequality in \eqref{eq:l-anticommutator}, gives
\begin{equation}\label{eq:L-ideal-bound}
\trnorm{L(\sigma)}
\le \trnorm{\sigma_{AR}\otimes\one_{A'}}
   \opnorm{S_{AA'}\otimes\one_R}.
\end{equation}
The swap $S_{AA'}$ is unitary, so $\opnorm{S_{AA'}\otimes\one_R}=1$. The trace norm is multiplicative under tensor products, and $\dim\HAp=d$, so
\begin{equation}\label{eq:identity-tensor-trace-norm}
\trnorm{\sigma_{AR}\otimes\one_{A'}}
=\trnorm{\sigma}\,\trnorm{\one_{A'}}
=d\,\trnorm{\sigma}.
\end{equation}
This proves the upper bound in (iii). 

(iv) is immediate because $S$ has only $0/1$ entries.
\end{proof}

\begin{theorem}[exact facial embedding]\label{thm:facial-embedding}
The set $F := \bQ(A{:}A'R)\cap\{X : \Tr[W_0X] = 0\}$ is an exposed face of $\bQ(A{:}A'R)$, and
\begin{equation}\label{eq:face-identity}
F \;=\; L\bigl(\SEPp(A{:}R)\bigr).
\end{equation}
Consequently, for every Hermitian $\sigma$ on $\HA\otimes\HR$,
\begin{equation}\label{eq:exact-reduction}
\sigma\in\SEPp(A{:}R)
\quad\Longleftrightarrow\quad
L(\sigma)\in\bQ(A{:}A'R).
\end{equation}
\end{theorem}

\begin{proof}
\emph{Exposed-face claim.} Use the definition from Section~\ref{sec:conventions} with $\ell(X):=\Tr[W_0X]$. Lemma~\ref{lem:witness} states that $\ell(X)\ge0$ for every $X\in\bQ(A{:}A'R)$. Moreover, $\bQ(A{:}A'R)$ is a cone, so it contains $0$, and $\ell(0)=0$. Thus the minimum of $\ell$ over $\bQ(A{:}A'R)$ is attained and equals $0$. Consequently,
\begin{equation}\label{eq:face-minimizers}
F=\{X\in\bQ(A{:}A'R):\ell(X)=0\}
=\operatorname*{arg\,min}_{X\in\bQ(A{:}A'R)}\ell(X),
\end{equation}
so $F$ is nonempty and exposed. Equivalently, for $d\ge2$, the hyperplane $\{X:\ell(X)=0\}$ supports the cone, and $F$ is its intersection with the cone. To verify the face property directly, suppose that $X_1,X_2\in\bQ(A{:}A'R)$, $0<t<1$, and $tX_1+(1-t)X_2\in F$. Then
\begin{equation}\label{eq:face-convex-combination}
0=t\ell(X_1)+(1-t)\ell(X_2).
\end{equation}
Both terms on the right are nonnegative by Lemma~\ref{lem:witness}, so $\ell(X_1)=\ell(X_2)=0$. Hence $X_1,X_2\in F$.

\emph{$L(\SEPp)\subseteq F$.} For a state $\tau$ on $\HR$, the map $\cE_\tau(X) := X\otimes\tau$ is CPTP, and by Lemma~\ref{lem:equality-case}, $Q_{\rho,\cE_\tau} = \Gamma(\rho)\otimes\tau = L(\rho\otimes\tau)$. Conic combinations give $L(\SEPp)\subseteq\bQ$; membership in $W_0^\perp$ is Lemma~\ref{lem:embedding-map}(ii).

\emph{$F\subseteq L(\SEPp)$.} The claim is immediate for $Y=0$. Let $0\ne Y\in F$. By Proposition~\ref{prop:cone-properties}(ii), $Y = \sum_\alpha t_\alpha Q_\alpha$, a finite sum with $t_\alpha>0$ and $Q_\alpha:=Q_{\rho_\alpha,\cE_\alpha}$. Then $0 = \Tr[W_0Y] = \sum_\alpha t_\alpha\Tr[W_0Q_\alpha]$ with every term nonnegative (Lemma~\ref{lem:witness}), so every term vanishes, and Lemma~\ref{lem:equality-case} gives $Q_\alpha = L(\rho_\alpha\otimes\tau_\alpha)$. Hence $Y = L(\omega)$ with $\omega = \sum_\alpha t_\alpha\,\rho_\alpha\otimes\tau_\alpha\in\SEPp$.

\emph{Equivalence.} ($\Rightarrow$) is $L(\SEPp)\subseteq\bQ$. ($\Leftarrow$): if $L(\sigma)\in\bQ$ then, since $L(\sigma)\perp W_0$ automatically (Lemma~\ref{lem:embedding-map}(ii)), $L(\sigma)\in F = L(\SEPp)$, so $L(\sigma) = L(\omega)$ for some $\omega\in\SEPp$; injectivity gives $\sigma = \omega\in\SEPp$.
\end{proof}

\begin{proposition}[stability of the face]\label{prop:stability}
Let $\eta := \Tr[W_0\,Q_{\rho,\cE}]$. Then there is $\tau\succeq0$ on $\HR$ with $\Tr[\tau]\le1$ such that $\Gamma(\rho)\otimes\tau\in L(\SEPp)$ and
\begin{equation}\label{eq:stability}
\trnorm{\,Q_{\rho,\cE} - \Gamma(\rho)\otimes\tau\,}
\;\le\; 2\sqrt2\; d^{3/2}\sqrt{\eta}.
\end{equation}
Consequently, for every $Y\in\bQ(A{:}A'R)$,
\begin{equation}\label{eq:stability-cone}
\dist_1\bigl(Y,\ L(\SEPp)\bigr)
\;:=\; \inf_{Z\in L(\SEPp)}\trnorm{Y-Z}
\;\le\; 2\sqrt2\; d^{3/2}\sqrt{\Tr[Y]\cdot\Tr[W_0Y]}.
\end{equation}
\end{proposition}

\begin{proof}
If $\eta\ge d$, take $\tau=0$. Then $\Gamma(\rho)\otimes\tau=0\in L(\SEPp)$ and $\trnorm{Q_{\rho,\cE}}\le\trnorm{J_\cE}\le d\,\trnorm{C_\cE}=d^2\le2\sqrt2\,d^{3/2}\sqrt\eta$. The norm inequalities used here are proved in Step 4 below.

Now assume $0\le\eta<d$. Fix Kraus operators $F_k\colon\HA\to\HAp\otimes\HR$ of $\cE$ and set $K_{ks}:=(\one_{A'}\otimes\bra{s})F_k$, a Kraus family for $\Lambda$ with joint index $(k,s)$. Let $a_{ks}:=\Tr[K_{ks}]$, let $b_{ks}:=\Tr[\rho K_{ks}]$, and decompose $K_{ks}=\tfrac{a_{ks}}d\one+\varepsilon_{ks}$ with $\Tr[\varepsilon_{ks}]=0$.

\emph{Step 1: the Kraus estimate.} From \eqref{eq:pairing} and \eqref{eq:b-bound}, $d - \eta = \operatorname{Re}\sum_{ks}\overline{a_{ks}}b_{ks} \le\abs{\sum_{ks}\overline{a_{ks}}b_{ks}}\le(\sum_{ks}\abs{a_{ks}}^2)^{1/2}$, so $\sum_{ks}\abs{a_{ks}}^2\ge(d-\eta)^2$, and therefore
\begin{equation}\label{eq:kraus-estimate}
\sum_{ks}\hsnorm{\varepsilon_{ks}}^2
= \sum_{ks}\Tr[K_{ks}^\dagger K_{ks}] - \frac1d\sum_{ks}\abs{a_{ks}}^2
= d - \frac1d\sum_{ks}\abs{a_{ks}}^2
\le d - \frac{(d-\eta)^2}{d} \le 2\eta .
\end{equation}

\emph{Step 2: the comparison map from the same Kraus family.} Define $\ket{\alpha_k} := \sum_s\tfrac{a_{ks}}{d}\ket{s}\in\HR$ and $F_k^0\ket{\psi} := \ket{\psi}\otimes\ket{\alpha_k}$. The completely positive map $\cE_0$ with Kraus family $\{F_k^0\}$ is $\cE_0(X) = X\otimes\tau$ with $\tau := \sum_k\ketbra{\alpha_k}{\alpha_k}\succeq0$ and $\Tr[\tau] = \sum_{ks}\abs{a_{ks}}^2/d^2 =: \lambda\le1$ by \eqref{eq:a-bound}. Its Jamio{\l}kowski matrix is $S\otimes\tau$, so $\tfrac12\{\rho\otimes\one, S\otimes\tau\} = \Gamma(\rho)\otimes\tau = L(\rho\otimes\tau)\in L(\SEPp)$.

\emph{Step 3: Choi comparison.} With $w_k := (\one\otimes F_k)\ket{\Omega}$ and $w_k^0 := (\one\otimes F_k^0)\ket{\Omega} = \ket{\Omega}\otimes\ket{\alpha_k}$ we have $C_\cE = \sum_kw_k^{\vphantom\dagger}w_k^\dagger$ and $C_{\cE_0} = \sum_kw_k^0(w_k^0)^\dagger$. Since $K_{ks}=(\one_{A'}\otimes\bra{s})F_k$, the operators $K_{ks}$ are the components of $F_k$ in the orthonormal basis $\{\ket{s}\}$ of $R$. The definitions in Steps 1 and 2 therefore give
\begin{equation}\label{eq:w-difference}
\begin{aligned}
F_k&=\sum_s K_{ks}\otimes\ket{s}_R,
&F_k^0&=\sum_s\frac{a_{ks}}d\one\otimes\ket{s}_R,\\
F_k-F_k^0&=\sum_s\varepsilon_{ks}\otimes\ket{s}_R,\\
\norm{w_k-w_k^0}^2
&=\hsnorm{F_k-F_k^0}^2
=\sum_s\hsnorm{\varepsilon_{ks}}^2,\\
\sum_k\norm{w_k}^2
&=\Tr\Bigl[\sum_kF_k^\dagger F_k\Bigr]=d,
&\sum_k\norm{w_k^0}^2
&=d\sum_k\norm{\alpha_k}^2=\lambda d\le d.
\end{aligned}
\end{equation}
The third line uses the vectorization isometry \eqref{eq:omega-isometry}. Its last equality holds because the terms $\varepsilon_{ks}\otimes\ket{s}_R$ are orthogonal in the Hilbert--Schmidt inner product for different values of $s$. The last line uses trace preservation, $\sum_kF_k^\dagger F_k=\one$, and the definition $\lambda=\Tr[\tau]=\sum_k\norm{\alpha_k}^2$. For vectors, $ww^\dagger - vv^\dagger = \tfrac12(w-v)(w+v)^\dagger + \tfrac12(w+v)(w-v)^\dagger$ gives $\trnorm{ww^\dagger - vv^\dagger}\le(\norm{w} + \norm{v})\norm{w - v}$. Summing and applying Cauchy--Schwarz with \eqref{eq:kraus-estimate} and \eqref{eq:w-difference},
\begin{equation}\label{eq:choi-difference}
\trnorm{C_\cE - C_{\cE_0}}
\le\Bigl(\sum_k(\norm{w_k}+\norm{w_k^0})^2\Bigr)^{1/2}
\Bigl(\sum_k\norm{w_k-w_k^0}^2\Bigr)^{1/2}
\le\sqrt{4d}\cdot\sqrt{2\eta} = 2\sqrt{2d\eta}.
\end{equation}
The first factor is at most $2\sqrt d$ because $(a+b)^2\le2a^2+2b^2$ and \eqref{eq:w-difference} bounds each of the two sums of squared vector norms by $d$. The second factor is at most $\sqrt{2\eta}$ by \eqref{eq:kraus-estimate}.

\emph{Step 4: transfer to the stote.} For a unit vector $z$ with Schmidt coefficients $s_i$, the partial-transpose formula $\trnorm{(zz^\dagger)^{\TA}}=(\sum_i s_i)^2$ is given in \cite[Prop.~8]{VidalWerner2002}. There are at most $d$ coefficients and $\sum_i s_i^2=1$, so Cauchy--Schwarz gives $(\sum_i s_i)^2\le d$. Spectral decomposition and the triangle inequality therefore give $\trnorm{Z^{\TA}}\le d\,\trnorm{Z}$ for every Hermitian $Z$. The trace-norm ideal property \cite[Eq.~(1.175)]{Watrous2018} also gives $\trnorm{\tfrac12\{\rho\otimes\one, M\}}\le\opnorm{\rho}\trnorm{M}\le\trnorm{M}$. Hence
\begin{equation}\label{eq:transfer}
\trnorm{Q_{\rho,\cE} - \Gamma(\rho)\otimes\tau}
= \bigl\lVert\tfrac12\bigl\{\rho\otimes\one,\ (C_\cE - C_{\cE_0})^{\TA}\bigr\}\bigr\rVert_1
\le d\cdot2\sqrt{2d\eta} = 2\sqrt2\,d^{3/2}\sqrt\eta .
\end{equation}

\emph{Cone version.} If $Y=0$, then \eqref{eq:stability-cone} holds because $0\in L(\SEPp)$. Now let $Y\ne0$. By Proposition~\ref{prop:cone-properties}(ii), there is a finite decomposition
\begin{equation}\label{eq:stability-conic-decomposition}
Y=\sum_\alpha t_\alpha Q_\alpha,
\qquad t_\alpha>0,
\end{equation}
where every $Q_\alpha=Q_{\rho_\alpha,\cE_\alpha}$ is a generator of $\bQ$. Set $\eta_\alpha:=\Tr[W_0Q_\alpha]\ge0$. For each $\alpha$, the generator bound \eqref{eq:stability} gives an operator
\begin{equation}\label{eq:stability-term-comparison}
Z_\alpha=\Gamma(\rho_\alpha)\otimes\tau_\alpha\in L(\SEPp)
\quad\text{such that}\quad
\trnorm{Q_\alpha-Z_\alpha}
\le 2\sqrt2\,d^{3/2}\sqrt{\eta_\alpha}.
\end{equation}
Define $Z:=\sum_\alpha t_\alpha Z_\alpha$. Since $L(\SEPp)$ is a convex cone, $Z\in L(\SEPp)$. Therefore, by the definition of the distance in \eqref{eq:stability-cone} and the triangle inequality,
\begin{equation}\label{eq:stability-conic-distance}
\begin{aligned}
\dist_1\bigl(Y,L(\SEPp)\bigr)
&\le \trnorm{Y-Z}\\
&\le \sum_\alpha t_\alpha\trnorm{Q_\alpha-Z_\alpha}\\
&\le 2\sqrt2\,d^{3/2}\sum_\alpha t_\alpha\sqrt{\eta_\alpha}.
\end{aligned}
\end{equation}
Cauchy--Schwarz gives
\begin{equation}\label{eq:stability-conic-cauchy-schwarz}
\sum_\alpha t_\alpha\sqrt{\eta_\alpha}
=\sum_\alpha\sqrt{t_\alpha}\sqrt{t_\alpha\eta_\alpha}
\le
\sqrt{\sum_\alpha t_\alpha}\,
\sqrt{\sum_\alpha t_\alpha\eta_\alpha}.
\end{equation}
Every generator has unit trace by Proposition~\ref{prop:cone-properties}(i). Hence $\Tr[Y]=\sum_\alpha t_\alpha$, while linearity gives $\Tr[W_0Y]=\sum_\alpha t_\alpha\eta_\alpha$. Substituting these two identities into the last two bounds proves \eqref{eq:stability-cone}.
\end{proof}

Numerical checks of the identities and bounds in this section are recorded in Appendix~\ref{app:checks}.

\section{Complexity and semidefinite non-representability}\label{sec:consequences}

The source problem for Theorems~\ref{thm:exact-hardness} and~\ref{thm:weak-hardness} is the following separability promise problem: given a rational state $\sigma$ on $\HA\otimes\HR$ ($\dim\HA=d$, $\dim\HR=r$), distinguish $\sigma\in\SEP(A{:}R)$ from $\dist_{\mathrm{HS}}(\sigma,\SEP)\ge\delta$, where $\delta=1/\mathrm{poly}(dr)$. This problem is strongly NP-hard by the separability results of Gurvits \cite{Gurvits2003} and Gharibian \cite{Gharibian2010}. Its YES class includes boundary separable states; it is therefore distinct from the two-sided weak-membership convention in Section~\ref{sec:geometry}. We call it the \emph{robust separability problem} below. Each of the two reductions applies the polynomial-time map $\sigma\mapsto L(\sigma)$ of Lemma~\ref{lem:embedding-map}, with an additional mixing step for weak membership. The target dimensions $d_A=d$ and $d_B=dr$ are polynomial in the source dimensions.

\begin{theorem}[exact membership is NP-hard]\label{thm:exact-hardness}
Deciding, given a rational Hermitian $X$ on $\HA\otimes\HAp\otimes\HR$, whether $X\in\bQ(A{:}A'R)$, is NP-hard.
\end{theorem}

\begin{proof}
Map the robust separability instance $\sigma$ to $X := L(\sigma)$, computable in polynomial time with rational entries (Lemma~\ref{lem:embedding-map}(iv)). By Theorem~\ref{thm:facial-embedding}, the equivalence is \emph{exact}: if $\sigma\in\SEP$ then $X\in\bQ$, and if $\sigma\notin\SEP$ --- in particular on every NO instance of the promise problem --- then $X\notin\bQ$. An algorithm for exact membership in $\bQ$ therefore decides the promise problem, which is NP-hard. No robustness analysis is needed.
\end{proof}

\begin{theorem}[weak membership is NP-hard]\label{thm:weak-hardness}
$\WMEM(\bQ(A{:}A'R))$ is NP-hard, with all accuracy parameters inverse-polynomially bounded (strong NP-hardness). Unless $\mathrm{P} = \mathrm{NP}$, there is no polynomial-time exact or weak membership test for $\bQ(A{:}A'R)$ or its dual, in the regime $d_A = d$, $d_B = dr$. In particular, their normalized bases admit no uniformly constructible polynomial-size rational SDP formulations satisfying the encoding and conditioning assumptions for polynomial-time weak optimization stated in the proof.
\end{theorem}

\paragraph{Proof sketch.}
The map $X=L(\sigma)$ preserves exact membership by Theorem~\ref{thm:facial-embedding}, but weak membership requires a quantitative separation from the boundary. For a YES instance, $X$ lies in the exposed face $F$ (Figure~\ref{fig:facial-embedding}). In the dimensions $d\ge2$ used in the reduction, $\Tr[W_0X]=0$ whereas $\Tr[W_0\one/N]=d-1/d>0$, so the trace-one slice of $F$ is a proper exposed face of $\base$. Thus $X$ lies on the boundary of $\base$ within the trace-one hyperplane, even if $\sigma$ is an interior point of $\SEP$. A weak-membership oracle can return either answer at such a boundary point, so exact membership alone does not suffice for the reduction.

We therefore replace $X$ by $X'=(1-\theta)X+\theta\one/N$, where $N=d^2r$ and $\theta>0$ is small, as illustrated in Figure~\ref{fig:mixing-step}. The maximally mixed state $\one/N$ is the centre of a Hilbert--Schmidt ball of radius $r_N=1/\sqrt{N(N-1)}$ contained in $\base$ (Lemma~\ref{lem:prepare-generators}; \cite{GurvitsBarnum2002}). For every YES instance, convexity then places the entire ball of radius $\theta r_N$ around $X'$ inside $\base$, with both balls taken in the trace-one hyperplane. Thus mixing provides the interior margin needed to force a YES answer at a sufficiently small oracle accuracy.

For a NO instance, Proposition~\ref{prop:stability} first gives a quantitative separation of $X$ from $\bQ$: if $X$ were too close to some $Y\in\bQ$, then $\Tr[W_0X]=0$ would force $\Tr[W_0Y]$ to be small, and the stability bound would place $Y$ close to the face $F=L(\SEPp)$. Taking a partial trace and normalizing would produce a separable state too close to $\sigma$, contradicting the NO promise; Figure~\ref{fig:facial-embedding} illustrates these comparisons. We choose $\theta$ so that mixing changes $X$ by at most half this trace-norm separation. Converting the remaining separation to Hilbert--Schmidt distance gives the required exterior margin for $X'$. Both the interior and exterior margins are inverse-polynomial, so one inverse-polynomial oracle accuracy distinguishes the two cases.

\begin{proof}
Set $N = d^2r$ and $r_N := 1/\sqrt{N(N-1)}$. Let $(\sigma,\delta)$ be an arbitrary instance of the robust separability problem defined at the start of this section, with $d,r\ge2$ and rational $0<\delta\le1$. Thus $\sigma$ is a rational state on $\HA\otimes\HR$, promised to satisfy either $\sigma\in\SEP(A{:}R)$ (a YES instance) or $\dist_{\mathrm{HS}}(\sigma,\SEP(A{:}R))\ge\delta$ (a NO instance). The following map is applied in both cases, without knowing which promise holds. Define
\begin{equation}\label{eq:instance-map}
X := L(\sigma), \qquad
X' := (1-\theta)\,X + \theta\,\frac{\one}{N},
\qquad
\theta := \frac{\varepsilon_*}{2(d+1)},
\qquad
\varepsilon_* := \frac{\delta^2}{1024\,d^3(d+1)} .
\end{equation}
Lemma~\ref{lem:embedding-map}(i), (iii), and (iv) imply that $X$ is Hermitian, $\Tr[X]=1$, $\trnorm{X}\le d$, and $X$ is computable in polynomial time with rational entries. The definition gives $0<\theta<1$, so $X'$ is Hermitian with $\Tr[X']=1$. The entries of $X'$ and the parameters $\theta$ and $\varepsilon_*$ are rational with bit size polynomial in the input length. The radius $r_N$ is used only in the estimates; it need not be encoded in the oracle query. All balls and interiors below are taken in the real affine space of Hermitian operators of trace one.

\emph{YES instances: an interior ball.} If $\sigma\in\SEP$, then $X\in\bQ\cap\{\Tr = 1\}=\base$ by Theorem~\ref{thm:facial-embedding}. For any traceless Hermitian $Z$ with $\hsnorm{Z}\le\theta r_N$,
\begin{equation}\label{eq:deep-point}
X' + Z = (1-\theta)X + \theta\Bigl(\frac{\one}{N} + \frac{Z}{\theta}\Bigr)\in\base
\end{equation}
by convexity: $\hsnorm{Z/\theta}\le r_N$, so $\one/N+Z/\theta$ lies in the Gurvits--Barnum ball contained in $\base$ (Lemma~\ref{lem:prepare-generators}; \cite{GurvitsBarnum2002}). Thus the Hilbert--Schmidt ball of radius $\theta r_N$ centred at $X'$ is contained in $\base$.

\emph{NO instances: separation from the cone.} Assume that $\dist_{\mathrm{HS}}(\sigma,\SEP)\ge\delta$. Suppose, for contradiction, that some $Y\in\bQ$ satisfies $\trnorm{X - Y}\le\varepsilon_2$ for an $\varepsilon_2$ with $0\le\varepsilon_2\le\varepsilon_*$. We will use the trace/operator-norm H\"older inequality $\abs{\Tr[AB]}\le\opnorm{A}\trnorm{B}$ for Hermitian $A,B$ \cite[Eq.~(1.174)]{Watrous2018}; in particular, $\abs{\Tr[B]}\le\trnorm{B}$ by taking $A=\one$. Also $\opnorm{W_0}\le d+1$, since $W_0=(d\one-S)\otimes\one_R$ and the swap $S$ has eigenvalues $\pm1$.
\begin{itemize}
\item Lemma~\ref{lem:embedding-map}(ii) gives $\Tr[W_0X]=0$, and Lemma~\ref{lem:witness} gives $\Tr[W_0Y]\ge0$. Consequently, $0\le\Tr[W_0Y]=\Tr[W_0(Y-X)]\le\opnorm{W_0}\trnorm{Y-X} \le(d+1)\varepsilon_2$.
\item Since $\Tr[X]=1$, the same inequality gives $\abs{\Tr[Y]-1}\le\trnorm{Y-X}\le\varepsilon_2$. In particular, $\Tr[Y]\le1+\varepsilon_2\le2$, because $\varepsilon_2\le\varepsilon_*\le1$.
\item Applying Proposition~\ref{prop:stability} with these two bounds gives an $\omega\in\SEPp$ such that
\begin{equation}\label{eq:no-side-stability}
\begin{aligned}
\trnorm{Y - L(\omega)}
&\le 2\sqrt2\,d^{3/2}\sqrt{2(d+1)\varepsilon_2} \;=:\; g,\\
g^2 &= 16\,d^3(d+1)\,\varepsilon_2 \le \frac{\delta^2}{64}.
\end{aligned}
\end{equation}
Here the distance in Proposition~\ref{prop:stability} is attained, since $L(\SEPp)=F$ is a nonempty closed set in a finite-dimensional space (Theorem~\ref{thm:facial-embedding} and Proposition~\ref{prop:cone-properties}). Thus $g\le\delta/8$. Since $\delta\le1$, the definition of $\varepsilon_*$ also gives the simpler bound $\varepsilon_2\le\varepsilon_*\le\delta/8$.
\item Taking the partial trace over $A'$ recovers $\sigma-\omega$ from $X-L(\omega)$ by Lemma~\ref{lem:embedding-map}(i). The partial trace is positive and trace preserving, so it contracts the trace norm \cite[Cor.~3.40]{Watrous2018}. Together with the triangle inequality, this gives $\trnorm{\sigma-\omega}\le\trnorm{X-L(\omega)} \le\trnorm{X-Y}+\trnorm{Y-L(\omega)}\le\varepsilon_2+g\le\delta/4$.
\item The operator $\omega$ is separable and positive, but its trace need not be one. To obtain a separable state, set $t:=\Tr[\omega]$. Then $\abs{t-1} = \abs{\Tr[\omega-\sigma]}\le\trnorm{\sigma-\omega}\le\delta/4$, so in particular $t\ge1-\delta/4\ge3/4>0$, using $\delta\le1$. The normalized state $\hat\omega:=\omega/t\in\SEP$ therefore obeys
\begin{equation}\label{eq:renormalize}
\begin{aligned}
\trnorm{\sigma - \hat\omega}
&\le \trnorm{\sigma-\omega} + \trnorm{\omega - \hat\omega}\\
&= \trnorm{\sigma-\omega} + \abs{1 - \tfrac1t}\,\trnorm{\omega}\\
&= \trnorm{\sigma-\omega} + \abs{t-1}\\
&\le 2\,\trnorm{\sigma-\omega} \le \frac{\delta}{2},
\end{aligned}
\end{equation}
using $\trnorm{\omega} = \Tr[\omega] = t$ for $\omega\succeq0$.
\end{itemize}
Since $\hsnorm{M}\le\trnorm{M}$ for every operator $M$ \cite[Eq.~(1.168)]{Watrous2018}, we conclude that
\begin{equation}\label{eq:source-promise-contradiction}
\dist_{\mathrm{HS}}(\sigma,\SEP)\le\hsnorm{\sigma-\hat\omega}
\le\trnorm{\sigma-\hat\omega}\le\delta/2<\delta.
\end{equation}
This contradicts the NO promise. Hence every $Y\in\bQ$ satisfies $\trnorm{X-Y}>\varepsilon_*$.

\emph{Preserving separation under mixing.} The perturbation satisfies $\trnorm{X-X'}=\theta\trnorm{X-\one/N}\le\theta(d+1)=\varepsilon_*/2$, using $\trnorm{X}\le d$ and $\trnorm{\one/N}=1$. For every $Y\in\bQ$, the triangle inequality therefore gives
\begin{equation}\label{eq:far-point}
\trnorm{X' - Y}
\;\ge\; \trnorm{X - Y} - \trnorm{X - X'}
\;>\; \varepsilon_* - \theta\,\trnorm{X - \one/N}
\;\ge\; \varepsilon_* - \theta(d+1)
\;=\; \frac{\varepsilon_*}{2},
\end{equation}
so mixing preserves at least half of the trace-norm separation.

To obtain the Hilbert--Schmidt separation required for weak membership, use $\trnorm{M}\le\sqrt{\operatorname{rank}(M)}\,\hsnorm{M} \le\sqrt N\,\hsnorm{M}$ for operators on the $N$-dimensional space \cite[Eq.~(1.170), p.~32]{Watrous2018}. Taking the infimum over $Y\in\bQ$ gives
\begin{equation}\label{eq:no-instance-hs-separation}
\dist_{\mathrm{HS}}(X',\bQ)
\ge \frac{\dist_1(X',\bQ)}{\sqrt N}
\ge \frac{\varepsilon_*}{2\sqrt N}.
\end{equation}
Since $\base\subseteq\bQ$, the same lower bound holds for $\dist_{\mathrm{HS}}(X',\base)$. Thus every NO instance of the robust separability problem maps to a point $X'$ at an inverse-polynomial Hilbert--Schmidt distance from $\base$. This is the separation needed for the NO case of the weak-membership reduction.

\emph{Choosing the oracle accuracy.} Since $r_N\ge1/N$ and $\sqrt N\le N$, choose the rational weak-membership accuracy
\begin{equation}\label{eq:weak-accuracy}
\varepsilon_W := \tfrac12\min\bigl(\theta/N,\ \varepsilon_*/(2N)\bigr)
=\frac{\varepsilon_*}{4N(d+1)},
\end{equation}
where the equality uses $\theta=\varepsilon_*/[2(d+1)]$. This choice satisfies $\varepsilon_W<\theta r_N$ and $\varepsilon_W<\varepsilon_*/(2\sqrt N)$. On a YES instance, the ball of radius $\theta r_N$ around $X'$ lies in $\base$. Thus $X'$ is in the $\varepsilon_W$-interior, and the oracle must assert that $X'$ is within $\varepsilon_W$ of $\base$. On a NO instance, \eqref{eq:no-instance-hs-separation} places $X'$ at distance greater than $\varepsilon_W$ from $\base$. The oracle must then assert that $X'$ is not in its $\varepsilon_W$-interior. These are the two alternatives in the definition of $\WMEM$ in Section~\ref{sec:geometry}; see also \cite[Def.~2.1.14]{GrotschelLovaszSchrijver1988}.

The two oracle outputs therefore distinguish the source promise classes. The map $(\sigma,\delta)\mapsto(X',\varepsilon_W)$ is computable in polynomial time. Moreover, $N=d^2r$ is polynomial in the source dimensions, and $\varepsilon_W=\delta^2/[4096\,N d^3(d+1)^2]$ remains inverse-polynomial when the source gap $\delta$ is inverse-polynomial. The strong NP-hardness of the source problem \cite[Thm.~1]{Gharibian2010} therefore proves the claimed strong NP-hardness of $\WMEM(\bQ)$. Consequently, a polynomial-time algorithm for this weak-membership problem would imply $\mathrm{P}=\mathrm{NP}$. The corresponding conclusion for the dual follows from Corollary~\ref{cor:optimization-dual}. An exact-membership algorithm would also solve weak membership: an exact YES answer satisfies the first weak-membership alternative, while an exact NO answer satisfies the second.

For the computational inference concerning SDP descriptions, polynomial size must be accompanied by the hypotheses needed for polynomial-time optimization. A sufficient set of hypotheses is a uniformly constructible semidefinite lift of the trace-one base with polynomially many variables and constraints, polynomial matrix size, and rational defining and projection data of polynomial bit size. The lifted feasible set must also have a supplied rational feasible centre and known positive rational inner and outer radius bounds, relative to its equality affine space, with polynomial bit size and polynomially bounded logarithmic radius ratio. Under these hypotheses, the Gr{\"o}tschel--Lov{\'a}sz--Schrijver SDP complexity theorem, as stated in \cite[Thm.~1.1]{deKlerkVallentin2016}, gives polynomial-time weak optimization. Proposition~\ref{prop:balls} and the oracle equivalences in \cite[Ch.~4]{GrotschelLovaszSchrijver1988} then give polynomial-time weak membership, contradicting the hardness result unless $\mathrm{P}=\mathrm{NP}$. Polynomial size alone does not supply these algorithmic hypotheses. The unconditional geometric obstruction to finite SDP representations is proved separately in Theorem~\ref{thm:no-lift}.
\end{proof}

\begin{remark}[scope of the hardness statements]\label{rem:hardness-scope}
The exact and weak membership problems for $\bQ(A{:}B)$, with variable input and output dimensions, are NP-hard. This follows because Theorems~\ref{thm:exact-hardness} and~\ref{thm:weak-hardness} establish NP-hardness on the subclass of instances in which $B$ is presented as $A'\otimes R$, with $\dim A'=\dim A$. The theorems do not establish hardness for every fixed pair of input and output dimensions. If $d_A$ divides $d_B$, then any output space of dimension $d_B$ can be identified unitarily with $A'\otimes R$ for $\dim A'=d_A$. Such a tensor-product identification is not available in general when $d_A$ does not divide $d_B$.
\end{remark}

\begin{corollary}[optimization and the dual cone]\label{cor:optimization-dual}
$\WOPT(\bQ(A{:}A'R))$ is NP-hard, and so are weak membership and weak optimization for the dual cone $\bQ(A{:}A'R)^*$.
\end{corollary}

\begin{proof}
By Proposition~\ref{prop:balls}, $\base$ is a well-bounded centered convex body, so the Gr{\"o}tschel--Lov{\'a}sz--Schrijver oracle equivalences apply \cite[Ch.~4]{GrotschelLovaszSchrijver1988}: for such a body, weak membership, weak separation, weak validity and weak optimization --- for the body \emph{and} for its polar --- are all polynomial-time oracle inter-reducible.

For completeness, set $c:=\one/N$ and $D:=\base-c$ in the trace-zero subspace, and let $D^\circ$ be its polar in that subspace. Normalize the nonzero elements of the dual cone by
\begin{equation}\label{eq:dual-base}
\base^*:=\{Z\in\bQ^*:\ip{Z}{c}=1\}.
\end{equation}
This normalization gives $\Tr[Z]=N$. Dividing by $N$ gives the trace-one base of the dual cone and rescales Hilbert--Schmidt accuracies by the known polynomial factor $N$.
The point $c$ is in the relative interior of $\base$, so $\ip{Z}{c}>0$ for every nonzero $Z\in\bQ^*$; thus $\base^*$ is a base of $\bQ^*$. Every $Z\in\base^*$ has the unique form $Z=\one+Z_0$ with $\Tr[Z_0]=0$, and
\begin{equation}\label{eq:dual-polar}
Z\in\bQ^*
\quad\Longleftrightarrow\quad
1+\ip{Z_0}{x}\ge0\ \text{for every }x\in D
\quad\Longleftrightarrow\quad
-Z_0\in D^\circ.
\end{equation}
Hence $\base^*=\one-D^\circ$. This is an explicit affine identification. The inner and outer ball bounds of Proposition~\ref{prop:balls} give polynomially related reciprocal bounds for $D^\circ$, so the identification has polynomial condition numbers. The oracle equivalences therefore transfer NP-hardness of $\WMEM(\bQ)$ (Theorem~\ref{thm:weak-hardness}) to $\WOPT(\bQ)$, $\WMEM(\bQ^*)$, and $\WOPT(\bQ^*)$.
\end{proof}

\begin{theorem}[no finite semidefinite representation]\label{thm:no-lift}
Assume $d,r\ge2$. If $d+r>5$, equivalently $dr>6$, then the cone $\bQ(A{:}A'R)$ with $\dim\HA=d$ and $\dim\HR=r$ is not a spectrahedral shadow: it admits no finite semidefinite-programming description of any size.
\end{theorem}

\begin{proof}
Suppose $\bQ = \pi(\mathcal{T})$ for a spectrahedron $\mathcal{T}\subseteq\mathbb{R}^m$ and a linear map $\pi$ \cite{BlekhermanParriloThomas2013}. The exposed face of Theorem~\ref{thm:facial-embedding} then satisfies
\begin{equation}\label{eq:face-of-shadow}
F = \bQ\cap W_0^\perp = \pi\bigl(\mathcal{T}\cap\pi^{-1}(W_0^\perp)\bigr):
\end{equation}
the preimage of the hyperplane $W_0^\perp$ is an affine subspace, whose intersection with $\mathcal{T}$ is again a spectrahedron, so $F$ is a spectrahedral shadow. By \eqref{eq:face-identity} and Lemma~\ref{lem:embedding-map}(i), $\SEPp(A{:}R) = (\Tr_{A'}\otimes\id_R)[F]$ is a linear image of $F$, hence also a spectrahedral shadow, and so is its trace-one affine slice $\SEP$. This contradicts Fawzi's theorem \cite{Fawzi2021}, which states that $\SEP(\mathbb{C}^d{:}\mathbb{C}^r)$ has no finite semidefinite representation when $d,r\ge2$ and $d+r>5$.
\end{proof}

\begin{remark}[upper bounds: the problem is semialgebraic]\label{rem:existential-R}
Membership in $\bQ$ is an existential semialgebraic condition. By Proposition~\ref{prop:cone-properties}(ii), it is equivalent to the existence of at most $N^2+1$ pairs $A_\alpha\succeq0$ on $A$ and $C_\alpha\succeq0$ on $AB$ such that $\Tr_B[C_\alpha]=\one_A$ and $X=\sum_\alpha(A_\alpha\otimes\one_B)\star C_\alpha^{\TA}$. The coefficient of each conic term is absorbed into the unnormalized input $A_\alpha$; conversely, a nonzero $A_\alpha$ has coefficient $\Tr[A_\alpha]$ and normalized input $A_\alpha/\Tr[A_\alpha]$. Write positivity as the Gram-factorization equations $A_\alpha=U_\alpha U_\alpha^\dagger$ and $C_\alpha=V_\alpha V_\alpha^\dagger$, retaining $A_\alpha$ and $C_\alpha$ as separate variables. After separating real and imaginary parts, these constraints and the reconstruction equation are quadratic equations in polynomially many real variables. Thus exact membership lies in $\exists\mathbb R$ and is decidable by real algebraic methods \cite{BochnakCosteRoy1998}. Dual non-membership also has an existential description: a state and channel whose stote has negative pairing with the proposed dual element. These upper bounds do not determine the exact complexity of either problem.
\end{remark}

\section{Open problems}\label{sec:discussion}

(1)~\emph{Dimensions.} The reduction needs $B = A'R$ with a full copy $A'$ of the input; the hard instances have $d_B = d_Ar$. What is the smallest output dimension supporting hardness --- in particular, is membership hard for $d_B$ constant or must $d_B$ grow with $d_A$?

(2)~\emph{Constant gap.} For explicitly given density matrices, separability at fixed positive Hilbert--Schmidt accuracy admits a randomized polynomial-time algorithm \cite{Malavolta2026}, improving the earlier quasipolynomial bound \cite{BrandaoChristandlYard2011}. Bernasconi and Malavolta \cite{BernasconiMalavolta2026} subsequently gave a deterministic polynomial-time algorithm, including multipartite separability at fixed accuracy. The present reduction maps a constant source gap to a target accuracy that decreases with dimension, so it does not imply constant-gap hardness for $\bQ$. Nor does a separability algorithm by itself give an algorithm for arbitrary elements of $\bQ$, since the map $L$ identifies only an exposed face. The complexity of weak membership in $\bQ$ at constant Hilbert--Schmidt accuracy remains open.

(3)~\emph{Dimension dependence of stability.} The exponent $\tfrac12$ in Proposition~\ref{prop:stability} is optimal even in fixed dimensions, as Remark~\ref{rem:stability-sharpness} shows. Can the factor $d^{3/2}$ or the numerical constant be improved? Such improvements would sharpen the accuracy bounds in Theorem~\ref{thm:weak-hardness}.

(4)~\emph{Membership tests through hierarchies.} Which structured families admit membership or non-membership certificates from low levels of inner and outer semidefinite approximations of $\bQ$? Membership in an inner approximation certifies membership in $\bQ$, while exclusion from an outer approximation certifies non-membership; otherwise the test is inconclusive. Even for a convergent hierarchy, the SDP size required at inverse-polynomial accuracy need not be polynomial in the dimension. Convergence alone therefore does not contradict the worst-case hardness in Theorem~\ref{thm:weak-hardness}. Nor does it imply an exact description at any finite level, so it does not contradict Theorem~\ref{thm:no-lift}. Quantitative convergence bounds would clarify the dependence of the computational cost on dimension and accuracy.

For separable states, Gharibian, Hecht and Rudolph \cite{GharibianHechtRudolph2026} prove quantitative lower bounds on the size of approximate SDP extended formulations in a specified formulation model, and on SDP-representable outer approximations within a prescribed trace distance. These results complement Fawzi's exact non-representability theorem \cite{Fawzi2021}. Can the facial embedding and stability bound established here yield corresponding quantitative lower bounds for approximations of $\bQ$? Such a transfer would require explicit control of normalization and approximation error, as well as the assumptions of the formulation model.

\section*{Author contribution and LLM use disclaimer}

All the one author contributed equally to this work. LLM assistance, in particular OpenAI's Codex, has been used to derive the results and to assist in writing the manuscript. The author takes full responsibility for the statements in the manuscript, including any errors.

\section*{Acknowledgements}

Thanks to John Calsamiglia and Andreas Winter for discussing the problem.

M.H.R. is supported by the National Research Foundation, Prime Minister's Office, Singapore, under its Campus for Research Excellence and Technological Enterprise (CREATE) program, Quantum Security and Resilience for Emerging Technologies Enterprise (QUASAR-CREATE).

\appendix
\section{Non-convexity, stability sharpness, and numerical checks}\label{app:checks}

At trace one, $\base$ is the convex hull of the stote generators. The following exact example shows that the generating set itself is not convex.

\begin{example}[an exact non-convexity witness at $(2,2)$]\label{ex:nonconvex-witness}
Let $\rho_1=\ketbra{0}{0}$ and $\rho_2=\ketbra{1}{1}$. The stote $Q_1:=Q_{\rho_1,\id}$ is formed from the input state $\rho_1$ and the identity channel. The stote $Q_2:=Q_{\rho_2,\cE_U}$ is formed from the input state $\rho_2$ and the unitary channel $\cE_U(X)=UXU^\dagger$, where $\sigma_x=\ketbra{0}{1}+\ketbra{1}{0}$ and
\begin{equation}\label{eq:clifford-unitary}
U = e^{i\frac{\pi}{3}\sigma_x}
= \tfrac12\begin{pmatrix} 1 & i\sqrt3 \\ i\sqrt3 & 1 \end{pmatrix}.
\end{equation}
We mix $Q_1$ and $Q_2$ with equal weights. Explicitly, $Q_1 = \ketbra{00}{00} + \tfrac12(\ketbra{01}{10} + \ketbra{10}{01})$ and, in the computational basis $\{\ket{00},\ket{01},\ket{10},\ket{11}\}$,
\begin{equation}\label{eq:Q2-matrix}
Q_2 = \frac18\begin{pmatrix}
0 & 0 & i\sqrt3 & 3 \\
0 & 0 & 1 & -i\sqrt3 \\
-i\sqrt3 & 1 & 6 & 2i\sqrt3 \\
3 & i\sqrt3 & -2i\sqrt3 & 2
\end{pmatrix},
\qquad
\omega := \tfrac12(Q_1 + Q_2).
\end{equation}
The mixture has marginal $\Tr_B[\omega] = \tfrac12\one_A$. Any channel producing it must have $J=2\omega$, since $(\tfrac12\one_A\otimes\one_B)\star J=J/2$, and hence must have a positive semidefinite Choi matrix $2\omega^{\TA}$ \cite{Choi1975,Watrous2018}. This candidate is
\begin{equation}\label{eq:inverted-choi}
2\,\omega^{\TA} = \frac18\begin{pmatrix}
8 & 0 & -i\sqrt3 & 5 \\
0 & 0 & 3 & i\sqrt3 \\
i\sqrt3 & 3 & 6 & 2i\sqrt3 \\
5 & -i\sqrt3 & -2i\sqrt3 & 2
\end{pmatrix},
\end{equation}
with spectrum $\tfrac14\{-1,\ 3-2\sqrt3,\ 3,\ 3+2\sqrt3\}$: the minimal eigenvalue is exactly $-\tfrac14$, so $\omega$ is not a single stote of its marginal, although both $Q_1, Q_2$ are stotes.
\end{example}

\begin{remark}[the stability exponent is optimal]\label{rem:stability-sharpness}
The exponent of the witness value $\eta$ in the stability bounds \eqref{eq:stability} and \eqref{eq:stability-cone} of Proposition~\ref{prop:stability} is optimal, as the following example shows. In the cone bound, this value is $\eta=\Tr[W_0Y]$.

Fix $d=2$ and any state $\tau$ on $R$. Let $U_t=\operatorname{diag}(e^{it},e^{-it})$, $\rho=\one_A/2$, and $\cE_t(X)=U_t XU_t^\dagger\otimes\tau$. Write $Q_t=Q_{\rho,\cE_t}$ and $V=S_{AA'}\otimes\one_R$. Lemma~\ref{lem:pairing} gives
\begin{equation}\label{eq:sharpness-witness-value}
\eta_t=\Tr[W_0Q_t]
=2-\frac{|\Tr[U_t]|^2}{2}
=2\sin^2t.
\end{equation}
Every $Y\in F=L(\SEPp)$ satisfies $VYV=Y$: this follows from $S\Gamma(M)S=\Gamma(M)$ and linearity. The map $P_-(Z)=(Z-VZV)/2$ is trace-norm contractive by the triangle inequality and unitary invariance. Since $Q_t=\tfrac12(\one\otimes U_t)S(\one\otimes U_t^\dagger)\otimes\tau$, direct calculation gives
\begin{equation}\label{eq:sharpness-antisymmetric-part}
P_-(Q_t)=\frac{i\sin(2t)}2
\bigl(\ketbra{10}{01}-\ketbra{01}{10}\bigr)\otimes\tau,
\qquad
\trnorm{P_-(Q_t)}=|\sin(2t)|.
\end{equation}
Consequently $\trnorm{Q_t-Y}\ge\trnorm{P_-(Q_t-Y)}=|\sin(2t)|$ for every $Y\in F$, and hence
\begin{equation}\label{eq:sharpness-distance-lower-bound}
\dist_1(Q_t,F)\ge|\sin(2t)|
=\sqrt{2\eta_t}\,|\cos t|.
\end{equation}
As $t\to0$, this rules out a uniform bound of the form $\dist_1(Q_t,F)\le C\eta_t^\alpha$ for any $\alpha>\tfrac12$, even if $C$ may depend on the fixed dimensions. In particular, a linear-in-$\eta$ bound cannot hold in general. Since $\Tr[Q_t]=1$, this also establishes sharpness of the exponent in the cone bound \eqref{eq:stability-cone}.
\end{remark}

\begin{samepage}
\begin{remark}[facial embedding]\label{rem:numerics}
The identities and bounds were verified numerically using NumPy~2.3.5. For each dimension pair $d,r\in\{2,3\}$, we tested $100$ random state--channel pairs with Kraus ranks from $1$ to $4$, $100$ random Hermitian inputs to $L$, and $100$ constructed equality-case channels. The checks covered the pairing formula, nonnegativity of the witness pairing, the left inverse of $L$, witness orthogonality, both trace-norm bounds, and the stability estimate using the comparison point constructed in the proof. We also tested $25$ four-term conic combinations per dimension pair. All identity residuals were below $3\times10^{-15}$; the largest ratio of the comparison-point distance to the stated bound was less than $0.269$ for generators and $0.130$ for conic combinations. These ratios are upper bounds on the corresponding distance-to-face ratios, not numerical minimizations over the face. We also checked the matrices and spectrum in Example~\ref{ex:nonconvex-witness}, verified the characteristic polynomial of eight times \eqref{eq:inverted-choi} using exact arithmetic, and checked the identities in Remark~\ref{rem:stability-sharpness} at three positive values of $t$. These checks are not used in the proofs.
\end{remark}
\end{samepage}



\begin{thebibliography}{HHP+17}
\small
\raggedright
\setlength{\itemsep}{2pt}
\setlength{\parsep}{0pt}

\bibitem[AT07]{AliprantisTourky2007}
Charalambos D. Aliprantis and Rabee Tourky.
\emph{Cones and Duality}, volume 84 of \emph{Graduate Studies in Mathematics}.
\href{https://doi.org/10.1090/gsm/084}{American Mathematical Society, 2007}.

\bibitem[ABD+24]{ArvidssonShukurEtAl2024}
David R. M. Arvidsson-Shukur, William F. Braasch, Jr., Stephan De Bi{\`e}vre, Justin Dressel, Andrew N. Jordan, Christopher Langrenez, Matteo Lostaglio, Jeff S. Lundeen, and Nicole Yunger Halpern.
``Properties and applications of the Kirkwood--Dirac distribution''.
\href{https://doi.org/10.1088/1367-2630/ada05d}{\emph{New J. Phys.} \textbf{26}, 121201 (2024)}.

\bibitem[BM26]{BernasconiMalavolta2026}
Martino Bernasconi and Giulio Malavolta.
``Polynomial-time algorithms for nuclear tensor norms and multipartite separability''.
\href{https://doi.org/10.48550/arXiv.2610.03637}{\emph{arXiv:2610.03637 [quant-ph]} (2026)}.

\bibitem[BPT13]{BlekhermanParriloThomas2013}
Grigoriy Blekherman, Pablo A. Parrilo, and Rekha R. Thomas, editors.
\emph{Semidefinite Optimization and Convex Algebraic Geometry}, volume 13 of \emph{MOS-SIAM Series on Optimization}.
\href{https://doi.org/10.1137/1.9781611972290}{SIAM, 2013}.

\bibitem[BCR98]{BochnakCosteRoy1998}
Jacek Bochnak, Michel Coste, and Marie-Fran{\c{c}}oise Roy.
\emph{Real Algebraic Geometry}, volume 36 of \emph{Ergebnisse der Mathematik und ihrer Grenzgebiete}.
\href{https://doi.org/10.1007/978-3-662-03718-8}{Springer, 1998}.

\bibitem[BCY11]{BrandaoChristandlYard2011}
Fernando G. S. L. Brand{\~a}o, Matthias Christandl, and Jon Yard.
``Faithful squashed entanglement''.
\href{https://doi.org/10.1007/s00220-011-1302-1}{\emph{Commun. Math. Phys.} \textbf{306}, 805--830 (2011)}.

\bibitem[Cho75]{Choi1975}
Man-Duen Choi.
``Completely positive linear maps on complex matrices''.
\href{https://doi.org/10.1016/0024-3795(75)90075-0}{\emph{Linear Algebra Appl.} \textbf{10}(3), 285--290 (1975)}.

\bibitem[dKV16]{deKlerkVallentin2016}
Etienne de Klerk and Frank Vallentin.
``On the Turing model complexity of interior point methods for semidefinite programming''.
\href{https://doi.org/10.1137/15M103114X}{\emph{SIAM J. Optim.} \textbf{26}(3), 1944--1961 (2016)}.

\bibitem[DT21]{DePalmaTrevisan2021}
Giacomo De Palma and Dario Trevisan.
``Quantum optimal transport with quantum channels''.
\href{https://doi.org/10.1007/s00023-021-01042-3}{\emph{Ann. Henri Poincar{\'e}} \textbf{22}, 3199--3234 (2021)}.

\bibitem[Dir45]{Dirac1945}
Paul A. M. Dirac.
``On the analogy between classical and quantum mechanics''.
\href{https://doi.org/10.1103/RevModPhys.17.195}{\emph{Rev. Mod. Phys.} \textbf{17}, 195--199 (1945)}.

\bibitem[Faw21]{Fawzi2021}
Hamza Fawzi.
``The set of separable states has no finite semidefinite representation except in dimension $3\times 2$''.
\href{https://doi.org/10.1007/s00220-021-04163-2}{\emph{Commun. Math. Phys.} \textbf{386}, 1319--1335 (2021)}.

\bibitem[FJV15]{FitzsimonsJonesVedral2015}
Joseph F. Fitzsimons, Jonathan A. Jones, and Vlatko Vedral.
``Quantum correlations which imply causation''.
\href{https://doi.org/10.1038/srep18281}{\emph{Sci. Rep.} \textbf{5}, 18281 (2015)}.

\bibitem[Ful25]{Fullwood2025}
James Fullwood.
``Quantum dynamics as a pseudo-density matrix''.
\href{https://doi.org/10.22331/q-2025-04-24-1719}{\emph{Quantum} \textbf{9}, 1719 (2025)}.

\bibitem[FP22]{FullwoodParzygnat2022}
James Fullwood and Arthur J. Parzygnat.
``On quantum states over time''.
\href{https://doi.org/10.1098/rspa.2022.0104}{\emph{Proc. R. Soc. A} \textbf{478}(2264), 20220104 (2022)}.

\bibitem[Gha10]{Gharibian2010}
Sevag Gharibian.
``Strong NP-hardness of the quantum separability problem''.
\href{https://doi.org/10.26421/QIC10.3-4-11}{\emph{Quantum Inf. Comput.} \textbf{10}(3--4), 343--360 (2010)}.

\bibitem[GHR26]{GharibianHechtRudolph2026}
Sevag Gharibian, Carsten Hecht, and Dorian Rudolph.
``Semidefinite extension complexity of the separable set, with applications to approximate disentanglers''.
\href{https://doi.org/10.48550/arXiv.2609.09033}{\emph{arXiv:2609.09033 [quant-ph]} (2026)}.

\bibitem[GMP16]{GolseMouhotPaul2016}
Fran{\c{c}}ois Golse, Cl{\'e}ment Mouhot, and Thierry Paul.
``On the mean field and classical limits of quantum mechanics''.
\href{https://doi.org/10.1007/s00220-015-2485-7}{\emph{Commun. Math. Phys.} \textbf{343}, 165--205 (2016)}.

\bibitem[GBB26]{GoswamiBarvarBera2026}
Kaumudibikash Goswami, Archana Barvar, and Manabendra Nath Bera.
``Multi-time states in quantum and operational probabilistic theories: An axiomatic framework''.
\href{https://doi.org/10.48550/arXiv.2610.22489}{\emph{arXiv:2610.22489 [quant-ph]} (2026)}.

\bibitem[GLS88]{GrotschelLovaszSchrijver1988}
Martin Gr{\"o}tschel, L{\'a}szl{\'o} Lov{\'a}sz, and Alexander Schrijver.
\emph{Geometric Algorithms and Combinatorial Optimization}, volume 2 of \emph{Algorithms and Combinatorics}.
\href{https://doi.org/10.1007/978-3-642-97881-4}{Springer, 1988}.

\bibitem[Gur03]{Gurvits2003}
Leonid Gurvits.
``Classical deterministic complexity of Edmonds' problem and quantum entanglement''.
\href{https://doi.org/10.1145/780542.780545}{In \emph{Proc. 35th Annual ACM Symposium on Theory of Computing (STOC '03)}, pages 10--19, 2003}.

\bibitem[GB02]{GurvitsBarnum2002}
Leonid Gurvits and Howard Barnum.
``Largest separable balls around the maximally mixed bipartite quantum state''.
\href{https://doi.org/10.1103/PhysRevA.66.062311}{\emph{Phys. Rev. A} \textbf{66}, 062311 (2002)}.

\bibitem[HCW26]{HoogstederCalsamigliaWinter2025}
Matt Hoogsteder-Riera, John Calsamiglia, and Andreas Winter.
``Approach to optimal quantum transport via states over time''.
\href{https://doi.org/10.22331/q-2026-10-14-2236}{\emph{Quantum} \textbf{10}, 2236 (2026)}.

\bibitem[HHP+17]{HorsmanHeunenPuseyBarrettSpekkens2017}
Dominic Horsman, Chris Heunen, Matthew F. Pusey, Jonathan Barrett, and Robert W. Spekkens.
``Can a quantum state over time resemble a quantum state at a single time?''.
\href{https://doi.org/10.1098/rspa.2017.0395}{\emph{Proc. R. Soc. A} \textbf{473}(2205), 20170395 (2017)}.

\bibitem[Jam72]{Jamiolkowski1972}
Andrzej Jamio{\l}kowski.
``Linear transformations which preserve trace and positive semidefiniteness of operators''.
\href{https://doi.org/10.1016/0034-4877(72)90011-0}{\emph{Rep. Math. Phys.} \textbf{3}(4), 275--278 (1972)}.

\bibitem[Kir33]{Kirkwood1933}
John G. Kirkwood.
``Quantum statistics of almost classical assemblies''.
\href{https://doi.org/10.1103/PhysRev.44.31}{\emph{Phys. Rev.} \textbf{44}, 31--37 (1933)}.

\bibitem[LS13]{LeiferSpekkens2013}
Matthew S. Leifer and Robert W. Spekkens.
``Towards a formulation of quantum theory as a causally neutral theory of Bayesian inference''.
\href{https://doi.org/10.1103/PhysRevA.88.052130}{\emph{Phys. Rev. A} \textbf{88}, 052130 (2013)}.

\bibitem[LN24]{LieNg2024}
Seok Hyung Lie and Nelly H. Y. Ng.
``Quantum state over time is unique''.
\href{https://doi.org/10.1103/PhysRevResearch.6.033144}{\emph{Phys. Rev. Research} \textbf{6}, 033144 (2024)}.

\bibitem[LBL+23]{LostaglioEtAl2023}
Matteo Lostaglio, Alessio Belenchia, Amikam Levy, Santiago Hern{\'a}ndez-G{\'o}mez, Nicole Fabbri, and Stefano Gherardini.
``Kirkwood--Dirac quasiprobability approach to the statistics of incompatible observables''.
\href{https://doi.org/10.22331/q-2023-10-09-1128}{\emph{Quantum} \textbf{7}, 1128 (2023)}.

\bibitem[Mal26]{Malavolta2026}
Giulio Malavolta.
``Quantum separability in polynomial time''.
\href{https://doi.org/10.48550/arXiv.2607.23773}{\emph{arXiv:2607.23773 [quant-ph]} (2026)}.

\bibitem[PB23]{ParzygnatBuscemi2023}
Arthur J. Parzygnat and Francesco Buscemi.
``Axioms for retrodiction: achieving time-reversal symmetry with a prior''.
\href{https://doi.org/10.22331/q-2023-05-23-1013}{\emph{Quantum} \textbf{7}, 1013 (2023)}.

\bibitem[PF23]{ParzygnatFullwood2023}
Arthur J. Parzygnat and James Fullwood.
``From time-reversal symmetry to quantum Bayes' rules''.
\href{https://doi.org/10.1103/PRXQuantum.4.020334}{\emph{PRX Quantum} \textbf{4}, 020334 (2023)}.

\bibitem[RAV+15]{RiedEtAl2015}
Katja Ried, Megan Agnew, Lydia Vermeyden, Dominik Janzing, Robert W. Spekkens, and Kevin J. Resch.
``A quantum advantage for inferring causal structure''.
\href{https://doi.org/10.1038/nphys3266}{\emph{Nat. Phys.} \textbf{11}, 414--420 (2015)}.

\bibitem[Roc70]{Rockafellar1970}
R. Tyrrell Rockafellar.
\emph{Convex Analysis}.
\href{https://doi.org/10.1515/9781400873173}{Princeton University Press, 1970}.

\bibitem[SP25]{SongParzygnat2025}
Minjeong Song and Arthur J. Parzygnat.
``Bipartite quantum states admitting a causal explanation''.
\href{https://doi.org/10.1116/5.0303300}{\emph{AVS Quantum Sci.} \textbf{7}, 045002 (2025)}.

\bibitem[VW02]{VidalWerner2002}
Guifr{\'e} Vidal and Reinhard F. Werner.
``Computable measure of entanglement''.
\href{https://doi.org/10.1103/PhysRevA.65.032314}{\emph{Phys. Rev. A} \textbf{65}, 032314 (2002)}.

\bibitem[Vil09]{Villani2009}
C{\'e}dric Villani.
\emph{Optimal Transport: Old and New}, volume 338 of \emph{Grundlehren der mathematischen Wissenschaften}.
\href{https://doi.org/10.1007/978-3-540-71050-9}{Springer, 2009}.

\bibitem[Wat18]{Watrous2018}
John Watrous.
\emph{The Theory of Quantum Information}.
\href{https://doi.org/10.1017/9781316848142}{Cambridge University Press, 2018}.

\bibitem[YN76]{YudinNemirovski1976}
David B. Yudin and Arkadi S. Nemirovski.
``Informational complexity and effective methods of solution for convex extremal problems''.
\emph{Ekonomika i Matematicheskie Metody} \textbf{12}, 357--369 (1976).

\end{thebibliography}
\end{document}